\documentclass[prl,superscriptaddress,twocolumn]{revtex4-2}
\usepackage{tikz}
\usepackage{graphics,graphicx, array, afterpage}
\usepackage{qcircuit,amsmath}
\usepackage{grffile}
\usepackage[export]{adjustbox}
\usepackage{lineno}
\usepackage{xcolor}
\usepackage{longtable}
\usepackage{braket}
\usepackage{array}
\usepackage{multirow}
\usepackage{enumerate}
\usepackage{amsmath}
\usepackage{amssymb}
\usepackage{esint}
\usepackage{amsthm}
\usepackage{times}
\usepackage{multirow}

\usepackage[colorlinks, citecolor=blue]{hyperref}
\hypersetup{linkcolor=magenta,urlcolor=blue,citecolor=blue,pdfstartview={FitH},urlcolor=blue}

\newcommand{\tr}[1]{\textnormal{Tr}[#1]}

\begin{document}
\title{Exact Hidden Markovian Dynamics in Quantum Circuits}

\author{He-Ran Wang}
\affiliation{Institute for Advanced Study, Tsinghua University, Beijing 100084,
People's Republic of China}

\author{Xiao-Yang Yang}
\affiliation{Institute for Advanced Study, Tsinghua University, Beijing 100084,
People's Republic of China}
\affiliation{Department of Physics, Tsinghua University, Beijing 100084, People's
Republic of China}

\author{Zhong Wang}
\email{wangzhongemail@tsinghua.edu.cn}
\affiliation{Institute for Advanced Study, Tsinghua University, Beijing 100084, People's Republic of China}

\begin{abstract}
Characterizing nonequilibrium dynamics in quantum many-body systems is a challenging frontier of physics. In this Letter, we systematically construct solvable nonintegrable quantum circuits that exhibit exact hidden Markovian subsystem dynamics. This feature thus enables accurately calculating local observables for arbitrary evolution time. Utilizing the influence matrix method, we show that the influence of the time-evolved global system on a finite subsystem can be analytically described by sequential, time-local quantum channels acting on the subsystem with an ancilla of finite Hilbert space dimension. The realization of exact hidden Markovian property is facilitated by a solvable condition on the underlying two-site gates in the quantum circuit. We further present several concrete examples with varying local Hilbert space dimensions to demonstrate our approach.
\end{abstract}

\maketitle
In isolated quantum many-body systems driven out of equilibrium, thermalization typically occurs, where local observables relax to their thermal-averaged expectation values after a finite time.
Heuristically, the global system serves as a thermal bath for the local subsystem \cite{Deutsch1991Quantum,Srednicki1994Chaos,Rigol2008Thermalization,Cazalilla2010Focus,Polkovnikov2011Noneq,Kosloff2013quantum}.
On the other hand, various counterexamples of thermalization have been extensively studied, including integrable models \cite{Sutherland2004beautiful}, many-body localization \cite{Nandkishore2015Manybody,Abanin2019Colloquium}, and quantum many-body scars \cite{Serbyn2021quantum,Moudgalya2022Quantum,Chandran2022Quantum}.
However, for both scenarios (following or violating thermalization), it poses a formidable challenge to accurately quantify the influence of the time-evolved macroscopic many-body system on its own subsystem, due to the exponentially large Hilbert space dimension in the thermodynamic limit and the quantum memory effects brought by the non-Markovianity. 

Recently, progress has been made in quantum circuits, where the unitary evolution is discretized to sequences of local unitary gates.
In particular, Refs. \cite{Banuls2009Matrix,Hermes2012Tensor,Hastings2015Connecting,Perez2022Light,Lerose2021Influence, Sonner2021Influence,Ye2021Constructing} have developed an efficient tensor-network approach to trace out the system, and encode the influence on the subsystem into the fixed point of the spatial transfer matrix, which is also known as the \textit{influence matrix} \cite{Lerose2021Influence, Sonner2021Influence}.
However, the intrinsic complexity of many-body dynamics typically leads to complicated influence matrices as the evolution time grows, restricting rigorous numerical and analytical treatment within this approach \cite{Foligno2023Temporal}.



Here, we introduce a novel approach to systematically constructing 1+1 D nonintegrable quantum circuits exhibiting exact hidden Markovian subsystem dynamics.
We introduce a solvable condition for the underlying unitary gates allowing for efficient contractions of quantum-circuit tensor networks for arbitrary evolution time.
We show that the time-evolved system can be traced out to a closed-form influence matrix in the matrix product state (MPS) representation of finite bond dimension for arbitrary evolution time, thus enabling numerical calculations of subsystem dynamics in an exact fashion.
Notably, we interpret the influence matrix as sequential quantum channels acting on the subsystem boundary. 
Hence, our work uncovers new principles leading to subsystem hidden Markovian property, and provides a promising testground to explore rich phenomena in quantum many-body dynamics through analytical tools.

\begin{figure*}[ht]
\includegraphics[width=1\linewidth]{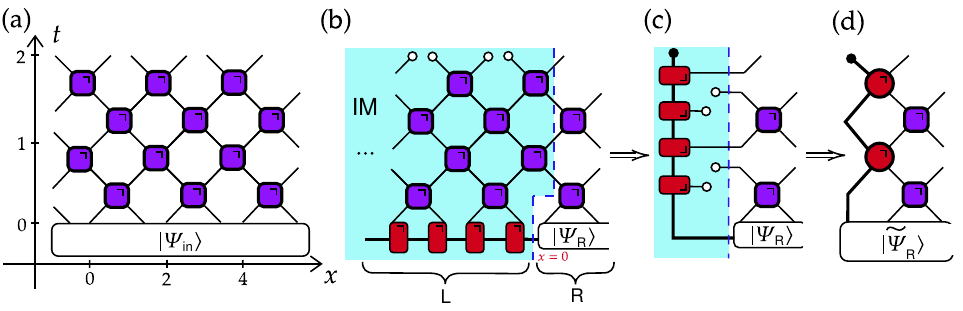} 
\caption{Main results of this letter. Total time steps $T=2$. (a) Tensor-network representation of a 1+1 D quantum circuit in the folded picture. The initial state $\ket{\Psi_{\text{in}}}$ is evolved by applying four layers of two-site gates [purple squares, defined in Eq. \eqref{eq:unitary_folded}] in a brickwork architecture.
(b) Illustration of the (left) influence matrix (IM). The system is initialized to a composite MPS over two regions: the left is a one-site shift invariant MPS [with local tensors in red, defined in Eq. \eqref{eq:tensor_folded}], and the right is a generic state. Thick lines correspond to the auxiliary Hilbert space. After attaching hollow dots on top outer legs in the left region, the tensor network in the light blue shaded region defines the influence matrix acting on the time slice (blue dotted line). 
(c) The exact influence matrix represented by MPS. 
(d) Open quantum system representation of the subsystem dynamics. Markovian property manifests when considering the joint dynamics of the ancilla in the auxiliary Hilbert space, and the subsystem. Two-site quantum channels are shown in red circles [defined in Eq. \eqref{eq:channel}].
}
\label{Illustration}
\end{figure*}

\textit{The setup}---In this letter we consider quantum circuits on a 1D lattice, where each site is labeled by an integer $x$. We associate a $q$-dimensional Hilbert space $\mathcal{H}_q$ for each site, with a basis: $\{\ket{a},a=0,1,\cdots,q-1\}$. 
The system is prepared in the initial state $\ket{\Psi_{\text{in}}}$ and undergoes discrete time evolution.
For each time step, the global unitary operator is
$\mathbb{U}=\mathbb{U}_{\text{odd}}\mathbb{U}_{\text{even}}$, where $\mathbb{U}_{\text{odd(even)}}=\otimes_{x\in {\text{odd(even)}}}U_{x,x+1}$. $U_{x,x+1}$ are two-site gates acting locally on $x$ and $x+1$.
As indicated by the form of $\mathbb{U}$, the local gates are arranged in a brickwork architecture. 

For the sake of convenience in depiction, we fold the forward and backward branches of time evolution [see Fig. \ref{Illustration}(a)]. By folding, each tensor is superimposed on its complex conjugate. The folded two-site unitary gate acting on the doubled space is defined as the following four-leg tensor:
\begin{equation}\label{eq:unitary_folded}
\tikzset{every picture/.style={line width=0.75pt}} 
\begin{tikzpicture}[x=0.75pt,y=0.75pt,yscale=-1,xscale=1]
\draw  [fill={rgb, 255:red, 144; green, 19; blue, 254 }  ,fill opacity=1 ][line width=1.5]  (141.5,183.96) .. controls (141.5,182.05) and (143.05,180.5) .. (144.96,180.5) -- (153.04,180.5) .. controls (154.95,180.5) and (156.5,182.05) .. (156.5,183.96) -- (156.5,192.04) .. controls (156.5,193.95) and (154.95,195.5) .. (153.04,195.5) -- (144.96,195.5) .. controls (143.05,195.5) and (141.5,193.95) .. (141.5,192.04) -- cycle ;
\draw  [fill={rgb, 255:red, 144; green, 19; blue, 254 }  ,fill opacity=1 ][line width=0.75]  (149,184) -- (153,184) -- (153,188) ;
\draw [fill={rgb, 255:red, 144; green, 19; blue, 254 }  ,fill opacity=1 ][line width=0.75]    (156.29,195.12) -- (166.5,205.5) ;
\draw [fill={rgb, 255:red, 144; green, 19; blue, 254 }  ,fill opacity=1 ][line width=0.75]    (131.5,170.5) -- (142,181) ;
\draw [fill={rgb, 255:red, 144; green, 19; blue, 254 }  ,fill opacity=1 ][line width=0.75]    (167.17,170.5) -- (156.28,181.44) ;
\draw [fill={rgb, 255:red, 144; green, 19; blue, 254 }  ,fill opacity=1 ][line width=0.75]    (142.17,194.17) -- (131.5,205.5) ;

\draw (189.17,175) node [anchor=north west][inner sep=0.75pt]    {$=U_{ab}^{cd}\left( U_{a'b'}^{c'd'}\right)^{*}$};
\draw (168,199.4) node [anchor=north west][inner sep=0.75pt]  [font=\scriptsize]  {$b,b'$};
\draw (110,161.4) node [anchor=north west][inner sep=0.75pt]  [font=\scriptsize]  {$c,c'$};
\draw (168,161.4) node [anchor=north west][inner sep=0.75pt]  [font=\scriptsize]  {$d,d'$};
\draw (110,199.4) node [anchor=north west][inner sep=0.75pt]  [font=\scriptsize]  {$a,a'$};
\end{tikzpicture},
\end{equation}
where $a,b,c,d$ ($a',b',c',d'$) denote the basis states in the forward (backward) branch. 

We focus on a certain class of initial states on the composite system $L$-$R$, with $L (R)$ for the left (right), as shown in Fig. \ref{Illustration}(b). 
For later convenience, we set the location of the leftmost site on R as the origin point $x=0$.
The global unitary operator admits a decomposition as $\mathbb{U}=\mathbb{U}_{\bar{R}}\mathbb{U}_{R}$, where $\mathbb{U}_{R}$ acts nontrivially on $R$, and $\mathbb{U}_{\bar{R}}$ acts on $L$ and the boundary across the two regions.
Meanwhile, we assume that the overall initial state can be decomposed as
$\ket{\Psi_{\text{in}}}=\sum_{j=1}^\chi\ket{\Psi_L^j}\otimes \ket{\Psi_R^j}$ (where each $\ket{\Psi_{R/L}^j}$ is not required to be normalized).
Here, we consider generic $\ket{\Psi_R^j}$ on the right, while $\ket{\Psi_L^j}$ can be written as a one-site shift-invariant MPS of bond dimension $\chi$ in terms of the three-leg tensor $A^{(a)}_{jk}$, where $j=0,1,\cdots,\chi-1$. 
The matrices $\{A^{(a)}\}$ act on the auxiliary Hilbert space $\mathcal{H}_\chi$ spanned by the basis $\{|j)\}_{j=0}^{\chi-1}$.
Graphically, we can represent the folded tensor $A$ as
\begin{equation}\label{eq:tensor_folded}
\tikzset{every picture/.style={line width=0.75pt}} 
\begin{tikzpicture}[x=0.75pt,y=0.75pt,yscale=-1,xscale=1]
\draw [line width=1.5]    (120,143.04) -- (149,143.04) ;
\draw  [fill={rgb, 255:red, 208; green, 2; blue, 27 }  ,fill opacity=1 ] (128.47,135.68) .. controls (128.47,134.42) and (129.49,133.4) .. (130.75,133.4) -- (137.59,133.4) .. controls (138.85,133.4) and (139.87,134.42) .. (139.87,135.68) -- (139.87,150.39) .. controls (139.87,151.65) and (138.85,152.67) .. (137.59,152.67) -- (130.75,152.67) .. controls (129.49,152.67) and (128.47,151.65) .. (128.47,150.39) -- cycle ;
\draw  [line width=0.75]  (133.3,135.63) -- (137.3,135.63) -- (137.3,139.63) ;

\draw [fill={rgb, 255:red, 144; green, 19; blue, 254 }  ,fill opacity=1 ][line width=0.75]    (134.2,124) -- (134.25,132.9) ;

\draw (124.8,112.21) node [anchor=north west][inner sep=0.75pt]  [font=\scriptsize]  {$a,a'$};
\draw (101.4,135.61) node [anchor=north west][inner sep=0.75pt]  [font=\scriptsize]  {$j,j'$};
\draw (150.4,136.61) node [anchor=north west][inner sep=0.75pt]  [font=\scriptsize]  {$k,k'$};
\draw (175,124.7) node [anchor=north west][inner sep=0.75pt]    {$=A_{jk}^{( a)} A{_{j'k'}^{( a')}}^{*}$};
\end{tikzpicture},
\end{equation}
such that the folded quantum state on $L$ has the form
\begin{equation*}
\tikzset{every picture/.style={line width=0.75pt}} 
\begin{tikzpicture}[x=0.75pt,y=0.75pt,yscale=-1,xscale=1]
\draw [fill={rgb, 255:red, 144; green, 19; blue, 254 }  ,fill opacity=1 ][line width=0.75]    (159.9,71.9) -- (159.95,80.8) ;
\draw [line width=1.5]    (230.98,91.64) -- (242.98,91.47) ;
\draw [line width=1.5]    (148,91.7) -- (249.57,91.7) ;
\draw  [fill={rgb, 255:red, 208; green, 2; blue, 27 }  ,fill opacity=1 ] (154.47,83.08) .. controls (154.47,81.82) and (155.49,80.8) .. (156.75,80.8) -- (163.59,80.8) .. controls (164.85,80.8) and (165.87,81.82) .. (165.87,83.08) -- (165.87,97.79) .. controls (165.87,99.05) and (164.85,100.07) .. (163.59,100.07) -- (156.75,100.07) .. controls (155.49,100.07) and (154.47,99.05) .. (154.47,97.79) -- cycle ;
\draw  [line width=0.75]  (159.3,83.03) -- (163.3,83.03) -- (163.3,87.03) ;

\draw  [fill={rgb, 255:red, 208; green, 2; blue, 27 }  ,fill opacity=1 ] (228.93,83.08) .. controls (228.93,81.82) and (229.95,80.8) .. (231.21,80.8) -- (238.05,80.8) .. controls (239.31,80.8) and (240.33,81.82) .. (240.33,83.08) -- (240.33,97.79) .. controls (240.33,99.05) and (239.31,100.07) .. (238.05,100.07) -- (231.21,100.07) .. controls (229.95,100.07) and (228.93,99.05) .. (228.93,97.79) -- cycle ;
\draw  [line width=0.75]  (233.77,83.03) -- (237.77,83.03) -- (237.77,87.03) ;

\draw  [fill={rgb, 255:red, 208; green, 2; blue, 27 }  ,fill opacity=1 ] (203.33,83.08) .. controls (203.33,81.82) and (204.35,80.8) .. (205.61,80.8) -- (212.45,80.8) .. controls (213.71,80.8) and (214.73,81.82) .. (214.73,83.08) -- (214.73,97.79) .. controls (214.73,99.05) and (213.71,100.07) .. (212.45,100.07) -- (205.61,100.07) .. controls (204.35,100.07) and (203.33,99.05) .. (203.33,97.79) -- cycle ;
\draw  [line width=0.75]  (208.17,83.03) -- (212.17,83.03) -- (212.17,87.03) ;

\draw  [fill={rgb, 255:red, 208; green, 2; blue, 27 }  ,fill opacity=1 ] (179.33,83.08) .. controls (179.33,81.82) and (180.35,80.8) .. (181.61,80.8) -- (188.45,80.8) .. controls (189.71,80.8) and (190.73,81.82) .. (190.73,83.08) -- (190.73,97.79) .. controls (190.73,99.05) and (189.71,100.07) .. (188.45,100.07) -- (181.61,100.07) .. controls (180.35,100.07) and (179.33,99.05) .. (179.33,97.79) -- cycle ;
\draw  [line width=0.75]  (184.17,83.03) -- (188.17,83.03) -- (188.17,87.03) ;

\draw [fill={rgb, 255:red, 144; green, 19; blue, 254 }  ,fill opacity=1 ][line width=0.75]    (234.3,71.9) -- (234.35,80.8) ;
\draw [fill={rgb, 255:red, 144; green, 19; blue, 254 }  ,fill opacity=1 ][line width=0.75]    (208.7,71.5) -- (208.75,80.4) ;
\draw [fill={rgb, 255:red, 144; green, 19; blue, 254 }  ,fill opacity=1 ][line width=0.75]    (185.1,71.9) -- (185.15,80.8) ;

\draw (140.97,91) node   [align=left] {\begin{minipage}[lt]{14.55pt}\setlength\topsep{0pt}
...
\end{minipage}};
\draw (252.47,87.87) node [anchor=north west][inner sep=0.75pt]  [font=\scriptsize]  {$j,j'$};
\draw (46.53,77.27) node [anchor=north west][inner sep=0.75pt]  [font=\normalsize]  {$|\Psi _{L}^{j} \rangle \langle \Psi _{L}^{j'} |=$};
\end{tikzpicture}.
\end{equation*}
We further assume that the MPS is in the left-canonical form \cite{Perez2007Matrix,Vidal2003Efficient,White2004Real} $\sum_{a=0}^{q-1}{A^{(a)}}^\dagger A^{(a)}=I_\chi$ such that the left boundary condition becomes unimportant in the thermodynamic limit.
Finally, we introduce the identity operators in the physical Hilbert space $\mathcal{H}_q$ and the auxiliary Hilbert space $\mathcal{H}_\chi$, represented by the hollow and solid dots, respectively,
\begin{equation}
\tikzset{every picture/.style={line width=0.75pt}}   
\begin{tikzpicture}[x=0.75pt,y=0.75pt,yscale=-1,xscale=1]
\draw [fill={rgb, 255:red, 144; green, 19; blue, 254 }  ,fill opacity=1 ][line width=0.75]    (174.9,85.5) -- (174.8,96.07) ;
\draw  [fill={rgb, 255:red, 255; green, 255; blue, 255 }  ,fill opacity=1 ] (172.1,82.7) .. controls (172.1,81.15) and (173.35,79.9) .. (174.9,79.9) .. controls (176.45,79.9) and (177.7,81.15) .. (177.7,82.7) .. controls (177.7,84.25) and (176.45,85.5) .. (174.9,85.5) .. controls (173.35,85.5) and (172.1,84.25) .. (172.1,82.7) -- cycle ;
\draw [fill={rgb, 255:red, 144; green, 19; blue, 254 }  ,fill opacity=1 ][line width=1.5]    (258.9,85.3) -- (258.8,95.87) ;
\draw  [fill={rgb, 255:red, 0; green, 0; blue, 0 }  ,fill opacity=1 ] (256.1,82.5) .. controls (256.1,80.95) and (257.35,79.7) .. (258.9,79.7) .. controls (260.45,79.7) and (261.7,80.95) .. (261.7,82.5) .. controls (261.7,84.05) and (260.45,85.3) .. (258.9,85.3) .. controls (257.35,85.3) and (256.1,84.05) .. (256.1,82.5) -- cycle ;

\draw (165.4,98.28) node [anchor=north west][inner sep=0.75pt]  [font=\footnotesize]  {$a,a'$};
\draw (183.6,82.8) node [anchor=north west][inner sep=0.75pt]    {$\ =\delta _{a,a'}$};
\draw (228,91) node [anchor=north west][inner sep=0.75pt]    {$,$};
\draw (249.4,98.08) node [anchor=north west][inner sep=0.75pt]  [font=\footnotesize]  {$j,j'$};
\draw (267.6,82.6) node [anchor=north west][inner sep=0.75pt]    {$\ =\delta _{j,j'} .$};
\end{tikzpicture}
\end{equation}

\textit{Exact influence matrix}---We would like to explore the influence of an infinitely large system on its own subsystem. 
To this end, in our setup we take the region $L$ to be semi-infinite. 
Next, we trace out the region $L$ after $T$ time steps of evolution, and thus obtain the reduced density matrix on $R$: $\rho_R(T)=\textnormal{Tr}_\text{L}[\mathbb{U}^T\ket{\Psi_{\text{in}}}\bra{\Psi_{\text{in}}}{\mathbb{U}^\dagger}^T]$. 
In Fig. \ref{Illustration}(b), the partial trace operation after time evolution is carried out by attaching hollow dots to the outer legs in $L$. 
Consequently, the evolution of the subsystem on $R$ can be expressed in terms of the internal dynamics $\mathbb{U}_R$, together with the action of a multitime operator accounting for the temporal correlations in the left bath. 
As shown in Fig. \ref{Illustration}(b), this operator lies on the multitime Hilbert space, which is obtained by tensoring those local Hilbert spaces carried by the legs cut by the time slice.
This operator is also referred to as the influence matrix (IM) \cite{Lerose2021Influence,Sonner2021Influence} or the process tensor \cite{Pollock2018non,Pollock2018Operational}.

We vectorize the IM to a quantum state in the doubled multitime Hilbert space \cite{Pollock2018non,Lerose2021Influence}.
Despite the class of dual unitary circuits and their generalizations that generate product-state IMs \cite{Bertini2019Exact,Piroli2020Exact,Jonay2021Triunitary,Yu2023Hierarchical,Bertini2023Exact,Liu2023Solvable,Foligno2024Quantum}, examples of analytically tractable IMs are limited to some exactly solvable models \cite{Klobas2021Exact,Lerose2021Scaling,Giudice2022Temporal}. 
On the other hand, in generic quantum circuits, the long-time IM usually becomes complicated, characterized by the bipartite entanglement entropy growing linearly with respect to the evolution time \cite{Lerose2021Influence,Sonner2021Influence,Foligno2023Temporal}.

Here, we find a solvable condition of the IM for arbitrarily long time. 
Notice that here the solvable condition serves as a sufficient but not necessary condition for obtaining closed-form influence matrices.
The condition is an algebraic relation only involving the local unitary gate and the local MPS,
\begin{equation}\label{eq:zipper}
\tikzset{every picture/.style={line width=0.75pt}}      
\begin{tikzpicture}[x=0.75pt,y=0.75pt,yscale=-1,xscale=1]
\draw  [fill={rgb, 255:red, 144; green, 19; blue, 254 }  ,fill opacity=1 ][line width=1.5]  (208.33,82.29) .. controls (208.33,80.38) and (209.88,78.83) .. (211.79,78.83) -- (219.87,78.83) .. controls (221.78,78.83) and (223.33,80.38) .. (223.33,82.29) -- (223.33,90.37) .. controls (223.33,92.28) and (221.78,93.83) .. (219.87,93.83) -- (211.79,93.83) .. controls (209.88,93.83) and (208.33,92.28) .. (208.33,90.37) -- cycle ;
\draw  [fill={rgb, 255:red, 144; green, 19; blue, 254 }  ,fill opacity=1 ][line width=0.75]  (215.83,82.33) -- (219.83,82.33) -- (219.83,86.33) ;
\draw [fill={rgb, 255:red, 144; green, 19; blue, 254 }  ,fill opacity=1 ][line width=0.75]    (223.13,93.45) -- (233.33,103.83) ;
\draw [fill={rgb, 255:red, 144; green, 19; blue, 254 }  ,fill opacity=1 ][line width=0.75]    (198.33,68.83) -- (208.83,79.33) ;
\draw [fill={rgb, 255:red, 144; green, 19; blue, 254 }  ,fill opacity=1 ][line width=0.75]    (234,68.83) -- (223.11,79.78) ;
\draw [fill={rgb, 255:red, 144; green, 19; blue, 254 }  ,fill opacity=1 ][line width=0.75]    (209,92.5) -- (198.33,103.83) ;
\draw  [fill={rgb, 255:red, 255; green, 255; blue, 255 }  ,fill opacity=1 ] (195.53,68.83) .. controls (195.53,67.29) and (196.79,66.03) .. (198.33,66.03) .. controls (199.88,66.03) and (201.13,67.29) .. (201.13,68.83) .. controls (201.13,70.38) and (199.88,71.63) .. (198.33,71.63) .. controls (196.79,71.63) and (195.53,70.38) .. (195.53,68.83) -- cycle ;
\draw [line width=1.5]    (188.5,61.94) -- (188.57,120.12) ;
\draw  [fill={rgb, 255:red, 208; green, 2; blue, 27 }  ,fill opacity=1 ] (195.89,100.33) .. controls (197.15,100.33) and (198.17,101.35) .. (198.17,102.61) -- (198.17,109.45) .. controls (198.17,110.71) and (197.15,111.73) .. (195.89,111.73) -- (181.18,111.73) .. controls (179.92,111.73) and (178.9,110.71) .. (178.9,109.45) -- (178.9,102.61) .. controls (178.9,101.35) and (179.92,100.33) .. (181.18,100.33) -- cycle ;
\draw  [line width=0.75]  (195.94,105.17) -- (195.94,109.17) -- (191.94,109.17) ;
\draw [line width=1.5]    (336.5,61.94) -- (336.57,120.12) ;
\draw  [fill={rgb, 255:red, 208; green, 2; blue, 27 }  ,fill opacity=1 ] (343.89,71.33) .. controls (345.15,71.33) and (346.17,72.35) .. (346.17,73.61) -- (346.17,80.45) .. controls (346.17,81.71) and (345.15,82.73) .. (343.89,82.73) -- (329.18,82.73) .. controls (327.92,82.73) and (326.9,81.71) .. (326.9,80.45) -- (326.9,73.61) .. controls (326.9,72.35) and (327.92,71.33) .. (329.18,71.33) -- cycle ;
\draw  [line width=0.75]  (343.94,76.17) -- (343.94,80.17) -- (339.94,80.17) ;
\draw    (346.5,76.13) -- (368,76.12) ;
\draw [fill={rgb, 255:red, 255; green, 255; blue, 255 }  ,fill opacity=1 ]   (346.33,103.83) -- (368.83,103.82) ;
\draw  [fill={rgb, 255:red, 255; green, 255; blue, 255 }  ,fill opacity=1 ] (343.53,103.83) .. controls (343.53,102.29) and (344.79,101.03) .. (346.33,101.03) .. controls (347.88,101.03) and (349.13,102.29) .. (349.13,103.83) .. controls (349.13,105.38) and (347.88,106.63) .. (346.33,106.63) .. controls (344.79,106.63) and (343.53,105.38) .. (343.53,103.83) -- cycle ;
\draw [fill={rgb, 255:red, 144; green, 19; blue, 254 }  ,fill opacity=1 ][line width=0.75]    (290.6,88.37) -- (305.07,103.47) ;
\draw [fill={rgb, 255:red, 144; green, 19; blue, 254 }  ,fill opacity=1 ][line width=0.75]    (273.33,68.86) -- (287.47,84.27) ;
\draw [fill={rgb, 255:red, 144; green, 19; blue, 254 }  ,fill opacity=1 ][line width=0.75]    (304.67,69.07) -- (273.33,103.86) ;
\draw  [fill={rgb, 255:red, 255; green, 255; blue, 255 }  ,fill opacity=1 ] (270.53,68.86) .. controls (270.53,67.31) and (271.79,66.06) .. (273.33,66.06) .. controls (274.88,66.06) and (276.13,67.31) .. (276.13,68.86) .. controls (276.13,70.41) and (274.88,71.66) .. (273.33,71.66) .. controls (271.79,71.66) and (270.53,70.41) .. (270.53,68.86) -- cycle ;
\draw [line width=1.5]    (263.5,61.97) -- (263.57,120.15) ;
\draw  [fill={rgb, 255:red, 208; green, 2; blue, 27 }  ,fill opacity=1 ] (270.89,100.36) .. controls (272.15,100.36) and (273.17,101.38) .. (273.17,102.64) -- (273.17,109.48) .. controls (273.17,110.74) and (272.15,111.76) .. (270.89,111.76) -- (256.18,111.76) .. controls (254.92,111.76) and (253.9,110.74) .. (253.9,109.48) -- (253.9,102.64) .. controls (253.9,101.38) and (254.92,100.36) .. (256.18,100.36) -- cycle ;
\draw  [line width=0.75]  (270.94,105.19) -- (270.94,109.19) -- (266.94,109.19) ;
\draw (236.8,88.14) node [anchor=north west][inner sep=0.75pt]    {$=$};
\draw (309,88.14) node [anchor=north west][inner sep=0.75pt]    {$=$};
\end{tikzpicture}.
\end{equation}
Conceptually, this condition implies that the unitary gate acts on the tensor $A$ as the {\scriptsize${\mathrm{SWAP}}$} gate in the spatial direction. 
Indeed, the solvable condition can be viewed as a refined version of zipper conditions \cite{Haegeman2017Diag}.
Previously, the zipper condition served as an ansatz to solve MPS influence matrices in integrable models \cite{Klobas2021Exact, Giudice2022Temporal}.
In contrast to those isolated examples, here we use the solvable condition as a criterion to construct generic nonintegrable quantum circuits with exact influence matrices \cite{zipper_footnote}.

In Fig. \ref{Illustration}(c), we directly present the exact form of the IM under the solvable condition (see  \cite{SuppMaterials} for detailed derivations).
An intuitive picture is that, by tracing out the left region, the initial MPS is rotated by $\pi/2$ and lies along the time slice, showing a novel manifestation of spacetime duality.
The IM is represented by a one-time-step shift-invariant MPS of bond dimension $\chi^2$ with appropriate boundary conditions. 
The replicated element in the MPS is the following four-leg tensor: 
\begin{equation}\label{eq:channel}
\tikzset{every picture/.style={line width=0.75pt}} 
\begin{tikzpicture}[x=0.75pt,y=0.75pt,yscale=-1,xscale=1]
\draw    (274.57,90.18) -- (283.07,90.22) ;
\draw [line width=1.5]    (248.82,51.39) -- (248.74,109.57) ;
\draw  [fill={rgb, 255:red, 208; green, 2; blue, 27 }  ,fill opacity=1 ] (255.94,60.77) .. controls (257.2,60.76) and (258.22,61.78) .. (258.22,63.04) -- (258.24,69.88) .. controls (258.24,71.14) and (257.22,72.16) .. (255.97,72.17) -- (241.26,72.2) .. controls (240,72.21) and (238.98,71.19) .. (238.97,69.93) -- (238.96,63.09) .. controls (238.95,61.83) and (239.97,60.81) .. (241.23,60.8) -- cycle ;
\draw  [line width=0.75]  (256,65.6) -- (256.01,69.6) -- (252.01,69.61) ;

\draw    (258.56,66.56) -- (283.06,66.61) ;
\draw [fill={rgb, 255:red, 255; green, 255; blue, 255 }  ,fill opacity=1 ]   (258.47,90.26) -- (264.97,90.18) ;
\draw  [fill={rgb, 255:red, 255; green, 255; blue, 255 }  ,fill opacity=1 ] (261.97,90.18) .. controls (261.96,88.63) and (263.21,87.38) .. (264.76,87.37) .. controls (266.31,87.37) and (267.56,88.62) .. (267.57,90.16) .. controls (267.57,91.71) and (266.32,92.97) .. (264.77,92.97) .. controls (263.23,92.98) and (261.97,91.72) .. (261.97,90.18) -- cycle ;
\draw  [fill={rgb, 255:red, 208; green, 2; blue, 27 }  ,fill opacity=1 ] (255.72,84.37) .. controls (256.98,84.36) and (258,85.38) .. (258,86.64) -- (258.02,93.48) .. controls (258.02,94.74) and (257,95.76) .. (255.74,95.77) -- (241.04,95.8) .. controls (239.78,95.81) and (238.76,94.79) .. (238.75,93.53) -- (238.74,86.69) .. controls (238.73,85.43) and (239.75,84.41) .. (241.01,84.4) -- cycle ;
\draw  [line width=0.75]  (255.78,89.2) -- (255.79,93.2) -- (251.79,93.21) ;

\draw  [fill={rgb, 255:red, 255; green, 255; blue, 255 }  ,fill opacity=1 ] (270.77,90.19) .. controls (270.76,88.64) and (272.01,87.38) .. (273.56,87.38) .. controls (275.11,87.38) and (276.36,88.63) .. (276.37,90.17) .. controls (276.37,91.72) and (275.12,92.98) .. (273.57,92.98) .. controls (272.03,92.98) and (270.77,91.73) .. (270.77,90.19) -- cycle ;
\draw  [color={rgb, 255:red, 0; green, 0; blue, 0 }  ,draw opacity=1 ][fill={rgb, 255:red, 208; green, 2; blue, 27 }  ,fill opacity=1 ][line width=1.5]  (183.2,79.42) .. controls (183.2,74.58) and (187.12,70.67) .. (191.95,70.67) .. controls (196.78,70.67) and (200.7,74.58) .. (200.7,79.42) .. controls (200.7,84.25) and (196.78,88.17) .. (191.95,88.17) .. controls (187.12,88.17) and (183.2,84.25) .. (183.2,79.42) -- cycle ;
\draw [fill={rgb, 255:red, 144; green, 19; blue, 254 }  ,fill opacity=1 ][line width=0.75]    (197.67,85.42) -- (208.48,95.53) ;
\draw [fill={rgb, 255:red, 144; green, 19; blue, 254 }  ,fill opacity=1 ][line width=0.75]    (210.64,61.57) -- (198.67,72.42) ;
\draw [line width=1.5]    (185.48,85.98) -- (170.05,103.98) ;
\draw [line width=1.5]    (168.12,57.55) -- (186.05,74.24) ;
\draw  [line width=0.75]  (191.9,75.37) -- (195.9,75.37) -- (195.9,79.37) ;

\draw (215.06,77.01) node [anchor=north west][inner sep=0.75pt]    {$=$};
\end{tikzpicture}.
\end{equation}
This local tensor can be represented explicitly in terms of $A$ \cite{SuppMaterials}, and replicates $T$ times in the IM.
At the lower boundary $t=0$, the auxiliary leg is connected to the initial state in the region $R$; at the upper boundary $t=T$, the auxiliary Hilbert space is traced out. 
Interestingly, the form of the IM indicates the nonsignaling property of the subsystem dynamics \cite{Piani2006Properties,Kofler2013Condition,Milz2019Completely,Hsieh2019Non}: the information flowing from the subsystem to the environment is completely discarded, as implied by the trace operator in Eq. \eqref{eq:channel}, and can never flow back into the subsystem. However, the environment retains the memory of initial correlations between $L$ and $R$.

\textit{Exact hidden Markovian subsystem dynamics}---Next, we show that the exact IM gives rise to hidden Markovian dynamics of the subsystem. 
As implied by the MPS representation, the IM is correlated in the time direction (when $\chi>1$), and thus the Markovian property is lacking \cite{Li2018Concepts,Pollock2018non,Pollock2018Operational,Milz2021Tutorial}.
Nevertheless, we can incorporate an ancilla of \textit{finite dimension} into the subsystem, which renders the joint dynamics to be governed by sequential quantum channels.
The ancilla lives in the auxiliary Hilbert space $\mathcal{H}_\chi$. 
We term such dynamics as \textit{hidden Markovian}, aligning with its classical counterpart \cite{Anderson1999Realization,Vidyasagar2011Complete,Kliesch2014Matrix}.
At $t=0$, we prepare the joint system in the pure state as $\ket{\tilde{\Psi}_R}=\sum_{j=0}^{\chi-1} |j)\bigotimes \ket{\Psi_R^j}$, which gives the same subsystem reduced density matrix when tracing out the ancilla. Here, $\bigotimes$ denotes tensor products between the auxiliary and physical Hilbert space.
Hence, the initial correlations between two regions are captured by this ancilla.
As illustrated in Fig. \ref{Illustration}(d), for each time step, the joint system is evolved by a layer of local gates, followed by a two-site quantum channel $\mathcal{M}$ acting on the ancilla and the leftmost site,
\begin{equation}\label{eq:full_time}
    \tilde{\rho}_R(t+1)=\mathcal{M}[\mathbb{U}_R \tilde{\rho}_R(t) \mathbb{U}_R^\dagger],
\end{equation}
where $\tilde{\rho}_R(t)$ is the joint system density matrix at time $t$. The dynamics of $\tilde{\rho}_R(t)$ is explicitly Markovian \footnote{This step can be traced back to the method of Markovian embedding, by incorporating additional degrees of freedom into the non-Markovian system to create the Markovian joint dynamics \cite{Imamoglu1994Stochastic,Garraway1997Non,Mazzola2009Pseudomodes,Smith2014Environmental,Kretschmer2016Collision,Campbell2018System,Tamascelli2018Non,Luchnikov2019Simulation,Mascherpa2020Optimized,Luchnikov2020Machine,Filippov2022Collisional,Flannigan2022Many}.}, where the state at the ($t+1$)th time step only depends on the state at $t$.
The quantum channel $\mathcal{M}$ is given by the four-leg tensor defined in Eq. \eqref{eq:channel}, which can be written in the standard Kraus form \cite{Nielsen2010Quantum,SuppMaterials}, $\mathcal{M}(\tilde{\rho}_R)=\sum_\mu K_\mu\tilde{\rho}_R K_\mu^\dagger$, where 
\begin{equation}\label{eq:Kraus}
    K_\mu=K_{a,a'}=\sum_{b=0}^{q-1} A^{(b)}A^{(a)}\bigotimes(\ket{b}\bra{a'})_{x=0}.
\end{equation}
Here $a,a'$ run over $0$ to $q-1$. Provided the Kraus-form representation, the quantum channel $\mathcal{M}$ is completely positive and trace-preserving \cite{Breuer2016Non}. The reduced density matrix on $R$ can be obtained by tracing out the ancilla at each time step: $\rho_R(t)=\text{Tr}_{\mathcal{H}_\chi}[\tilde{\rho}_R(t)]$.

A few remarks are in order.
First, we can prepare the left initial state in a matrix product density operator, instead of a MPS pure state. All the analysis works as well, except that the expression of Kraus form should be slightly modified \cite{SuppMaterials}. 
Second, the left initial states can be extended to a certain class of two-site shift-invariant MPS, while keeping the solvability of the influence matrix \cite{SuppMaterials}.
Third, due to the chiral structure in the solvable condition, it only allows efficient contractions of tensor networks from left to right, but not vice versa. 
However, by imposing an additional condition as follows,
\begin{equation}
\tikzset{every picture/.style={line width=0.75pt}} 
\begin{tikzpicture}[x=0.75pt,y=0.75pt,yscale=-1,xscale=1]
\draw  [fill={rgb, 255:red, 144; green, 19; blue, 254 }  ,fill opacity=1 ][line width=1.5]  (208.33,82.29) .. controls (208.33,80.38) and (209.88,78.83) .. (211.79,78.83) -- (219.87,78.83) .. controls (221.78,78.83) and (223.33,80.38) .. (223.33,82.29) -- (223.33,90.37) .. controls (223.33,92.28) and (221.78,93.83) .. (219.87,93.83) -- (211.79,93.83) .. controls (209.88,93.83) and (208.33,92.28) .. (208.33,90.37) -- cycle ;
\draw  [fill={rgb, 255:red, 144; green, 19; blue, 254 }  ,fill opacity=1 ][line width=0.75]  (215.83,82.33) -- (219.83,82.33) -- (219.83,86.33) ;
\draw [fill={rgb, 255:red, 144; green, 19; blue, 254 }  ,fill opacity=1 ][line width=0.75]    (223.13,93.45) -- (233.33,103.83) ;
\draw [fill={rgb, 255:red, 144; green, 19; blue, 254 }  ,fill opacity=1 ][line width=0.75]    (198.33,68.83) -- (208.83,79.33) ;
\draw [fill={rgb, 255:red, 144; green, 19; blue, 254 }  ,fill opacity=1 ][line width=0.75]    (234,68.83) -- (223.11,79.78) ;
\draw [fill={rgb, 255:red, 144; green, 19; blue, 254 }  ,fill opacity=1 ][line width=0.75]    (209,92.5) -- (198.33,103.83) ;

\draw  [fill={rgb, 255:red, 255; green, 255; blue, 255 }  ,fill opacity=1 ] (231.2,68.83) .. controls (231.2,67.29) and (232.45,66.03) .. (234,66.03) .. controls (235.55,66.03) and (236.8,67.29) .. (236.8,68.83) .. controls (236.8,70.38) and (235.55,71.63) .. (234,71.63) .. controls (232.45,71.63) and (231.2,70.38) .. (231.2,68.83) -- cycle ;
\draw [line width=1.5]    (243.5,59.94) -- (243.57,119.12) ;
\draw  [fill={rgb, 255:red, 208; green, 2; blue, 27 }  ,fill opacity=1 ] (235.18,110.73) .. controls (233.92,110.73) and (232.9,109.71) .. (232.9,108.45) -- (232.9,101.61) .. controls (232.9,100.35) and (233.92,99.33) .. (235.18,99.33) -- (249.89,99.33) .. controls (251.15,99.33) and (252.17,100.35) .. (252.17,101.61) -- (252.17,108.45) .. controls (252.17,109.71) and (251.15,110.73) .. (249.89,110.73) -- cycle ;
\draw  [line width=0.75]  (235.13,105.9) -- (235.13,101.9) -- (239.13,101.9) ;

\draw [line width=1.5]    (375.6,120.05) -- (375.6,61.87) ;
\draw  [fill={rgb, 255:red, 208; green, 2; blue, 27 }  ,fill opacity=1 ] (368.15,80.71) .. controls (366.89,80.72) and (365.86,79.7) .. (365.85,78.44) -- (365.81,71.6) .. controls (365.8,70.34) and (366.81,69.32) .. (368.07,69.31) -- (382.78,69.21) .. controls (384.04,69.2) and (385.06,70.21) .. (385.07,71.47) -- (385.12,78.31) .. controls (385.13,79.57) and (384.12,80.6) .. (382.86,80.61) -- cycle ;
\draw  [line width=0.75]  (368.07,75.88) -- (368.04,71.88) -- (372.04,71.85) ;

\draw    (365.5,75.1) -- (345,75.09) ;
\draw [fill={rgb, 255:red, 255; green, 255; blue, 255 }  ,fill opacity=1 ]   (367.63,104.3) -- (345.13,104.31) ;
\draw  [fill={rgb, 255:red, 255; green, 255; blue, 255 }  ,fill opacity=1 ] (370.43,104.57) .. controls (370.42,106.12) and (369.15,107.36) .. (367.6,107.34) .. controls (366.06,107.32) and (364.82,106.06) .. (364.83,104.51) .. controls (364.85,102.96) and (366.12,101.72) .. (367.66,101.74) .. controls (369.21,101.76) and (370.45,103.02) .. (370.43,104.57) -- cycle ;
\draw [fill={rgb, 255:red, 144; green, 19; blue, 254 }  ,fill opacity=1 ][line width=0.75]    (288.06,87.5) -- (274.33,102.69) ;
\draw [fill={rgb, 255:red, 144; green, 19; blue, 254 }  ,fill opacity=1 ][line width=0.75]    (304.45,67.9) -- (291.04,83.39) ;
\draw [fill={rgb, 255:red, 144; green, 19; blue, 254 }  ,fill opacity=1 ][line width=0.75]    (274.71,68.11) -- (304.45,103.08) ;
\draw  [fill={rgb, 255:red, 255; green, 255; blue, 255 }  ,fill opacity=1 ] (307.11,67.9) .. controls (307.11,66.34) and (305.92,65.08) .. (304.45,65.08) .. controls (302.99,65.08) and (301.8,66.34) .. (301.8,67.9) .. controls (301.8,69.45) and (302.99,70.71) .. (304.45,70.71) .. controls (305.92,70.71) and (307.11,69.45) .. (307.11,67.9) -- cycle ;
\draw [line width=1.5]    (313.79,60.97) -- (313.72,119.45) ;
\draw  [fill={rgb, 255:red, 208; green, 2; blue, 27 }  ,fill opacity=1 ] (306.9,99.56) .. controls (305.64,99.56) and (304.61,100.59) .. (304.61,101.85) -- (304.61,108.73) .. controls (304.61,109.99) and (305.64,111.02) .. (306.9,111.02) -- (321.49,111.02) .. controls (322.75,111.02) and (323.78,109.99) .. (323.78,108.73) -- (323.78,101.85) .. controls (323.78,100.59) and (322.75,99.56) .. (321.49,99.56) -- cycle ;
\draw  [line width=0.75]  (310.34,101.86) -- (306.61,101.85) -- (306.6,105.94) ;

\draw (323.8,83.14) node [anchor=north west][inner sep=0.75pt]    {$=$};
\draw (253,82.74) node [anchor=north west][inner sep=0.75pt]    {$=$};
\end{tikzpicture},
\end{equation}
contractions from the right are allowed as well (see \cite{SuppMaterials} for concrete examples). Quantum circuits equipped with two solvable conditions of opposite chirality facilitate analytical descriptions for more physical quantities, e.g., R$\acute{\text{e}}$nyi entropy dynamics \cite{Bertini2022Growth}. Specifically, in \cite{SuppMaterials} we show that the $n$th entanglement velocity $v_E^{(n)}$ \cite{Zhou2022Maximal} -- the asymptotic growing rate of the $n$th R$\acute{\text{e}}$nyi entropy ($n>1$) between two regions -- is given by the leading eigenvalue $\lambda_n$ of a certain quantum channel,
\begin{equation}\label{eq:Renyi}
    v_E^{(n)}\equiv\lim_{t\to\infty}\frac{\ln(\tr{\rho_R^n(t)})/(1-n)}{t\ln(q)}=\frac{2\ln(\lambda_n)}{(1-n)\ln(q)}.
\end{equation}
The entanglement velocity explicitly depends on the R$\acute{\text{e}}$nyi index $n$, which is clearly different from dual-unitary gates with solvable initial state \cite{Piroli2020Exact}.

Next, we demonstrate that solutions (up to one-site unitaries) of Eq. \eqref{eq:zipper} actually depend only on two characteristic values, $q$ and $\tilde{q}$, of the local tensor $A^{(a)}_{jk}$, rather than on exact values of all the tensor elements. Here, $q$ is the local Hilbert space dimension, and $\tilde{q}$ is the one-site MPS dimension which will be clarified later. For convenience, we define the reshuffled operator $U^R$ as $(U^R)_{ab}^{cd}=U_{ac}^{bd}$.
When read along the space direction, the solvable condition can be expressed as 
\begin{equation}
U^R(\ket{A_{jk}}\bra{A_{j'k'}}\otimes I_q)(U^R)^\dagger=I_q\otimes\ket{A_{jk}}\bra{A_{j'k'}},
\end{equation}
for all combinations of $j,j',k,k'$. Here, $I_q$ is the identity operator in the local Hilbert space, and the one-site MPS is defined as $\ket{A_{jk}}=\sum_{a=0}^{q-1}A^{(a)}_{jk}\ket{a}$.
This expression suggests that the set of solutions $U$ essentially depends on the Hilbert subspace $\mathcal{H}_A$ spanned by $\{\ket{A_{jk}}\}_{j,k=1}^{\chi}$. 
Furthermore, due to the isomorphism between different Hilbert subspaces with the same dimension, the solutions should solely rely on the subspace dimension $\tilde{q}=\text{dim}(\mathcal{H}_A)$, up to one-site unitaries.

Since the set of solutions applies to all initial states with the same $q$ and $\tilde{q}$, we will label the solutions by these two characteristic values in the following.


\textit{Example I}---In this and the next part, we will show that the solvable condition indeed leads to a wide range of solutions for quantum circuits. First, we present the solutions of $q=2, \tilde{q}=1$. For the reason demonstrated above, we can take a specific left initial state $\otimes_{x<0}\ket{0}_x$. In the context of MPS, the tensors are given by $A^{(0)}=1,A^{(1)}=0$, and thus $\mathcal{H}_A=\{\ket{0}\}$.
We provide an exhaustive parametrization for the solutions \cite{SuppMaterials}
\begin{equation}\label{eq:sol_Eq1}
    U=e^{i\phi}(u\otimes e^{-i\epsilon\sigma^3})V[J](e^{-i\eta\sigma^3}\otimes v),
\end{equation}
where $u,v\in \text{SU}(2)$, and
\begin{equation*}
    V[J]=\exp{[-i(\frac{\pi}{4}\sigma^1\otimes\sigma^1+\frac{\pi}{4}\sigma^2\otimes\sigma^2+J\sigma^3\otimes\sigma^3)]}.
\end{equation*}
Here, $\sigma^\alpha$, $\alpha=1,2,3$ are standard Pauli matrices. This class of solutions coincides with those two-qubit circuits featured in chiral solitons, which are dual unitary \cite{Bertini2020OperatorII}.

Compared to those previously known solvable initial states for dual-unitary circuits \cite{Bertini2019Exact}, our findings suggest a new initial state ($\otimes_{x<0}\ket{0}_x$) allowing for exact influence matrix in the product-state form.
The corresponding one-site quantum channel can be obtained by substituting the form of $A$ into Eq. \eqref{eq:Kraus}: $\mathcal{M}[\rho_R]=(\ket{0}\bra{0})_{x=0}\otimes\text{Tr}_{x=0}[\rho_R]$. 
Hence, the left bath acts as the boundary resetting toward $\ket{0}$, and thus the entanglement between two regions never grows up. 
In addition, we point out that for the solutions Eq. \eqref{eq:sol_Eq1},  we can independently choose the one-site initial state as $\ket{0}$ or $\ket{1}$ while holding the solvability of the influence matrix.

\textit{Example II}---Next, we present a (nonexhaustive) parameterization for solutions with $q=4,\tilde{q}=2$, where $\mathcal{H}_A$ is spanned by $\ket{0}$ and $\ket{1}$,
\begin{equation}\label{eq:q4chi2}
    U=e^{i\phi}W_2SW_1(I\otimes v).
\end{equation}
Here, $v\in \text{SU}(4)$, $S$ is the {\scriptsize${\mathrm{SWAP}}$} gate, $W_{1,2}$ are one-site controlled gates,
\begin{equation}
    W_{1,2}=\sum_{a=0}^{3} f_{1,2}^{(a)}\otimes \ket{a}\bra{a},
\end{equation}
where $f_1^{(a)}=
\left( \begin{array}{cc}
I_2 & 0 \\
0 & g^{(a)} 
\end{array} \right)$, $g^{(a)}\in\text{SU}(2)$, and $f_{2}^{(a)}\in \text{SU}(4)$ for all $a$. 
The parametrization can be generalized to higher $q$ and $\tilde{q}$ with slight modifications: $v\in\text{SU}(q)$, $f_1^{(a)}=I_{\tilde{q}}\oplus g^{(a)}$, where $g^{(a)}\in \text{SU}(q-\tilde{q})$, and $f_2^{(a)}\in\text{SU}(q)$. Generally, these solutions form an overlapping but different set with dual-unitary circuits. An exception is when $q-\tilde{q}=0$ or $1$, where $f_1^{(a)}=I_q,W_1=I_{q^2}$ and thus Eq. \eqref{eq:q4chi2} is reduced to a subclass of dual-unitary gates \cite{Borsi2022Construction,Prosen2021Many}.

\begin{figure}
\includegraphics[width=0.45\linewidth]{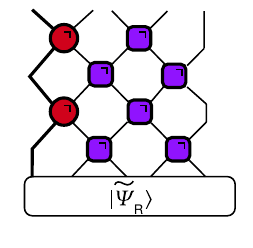} 
\includegraphics[width=0.5\linewidth]{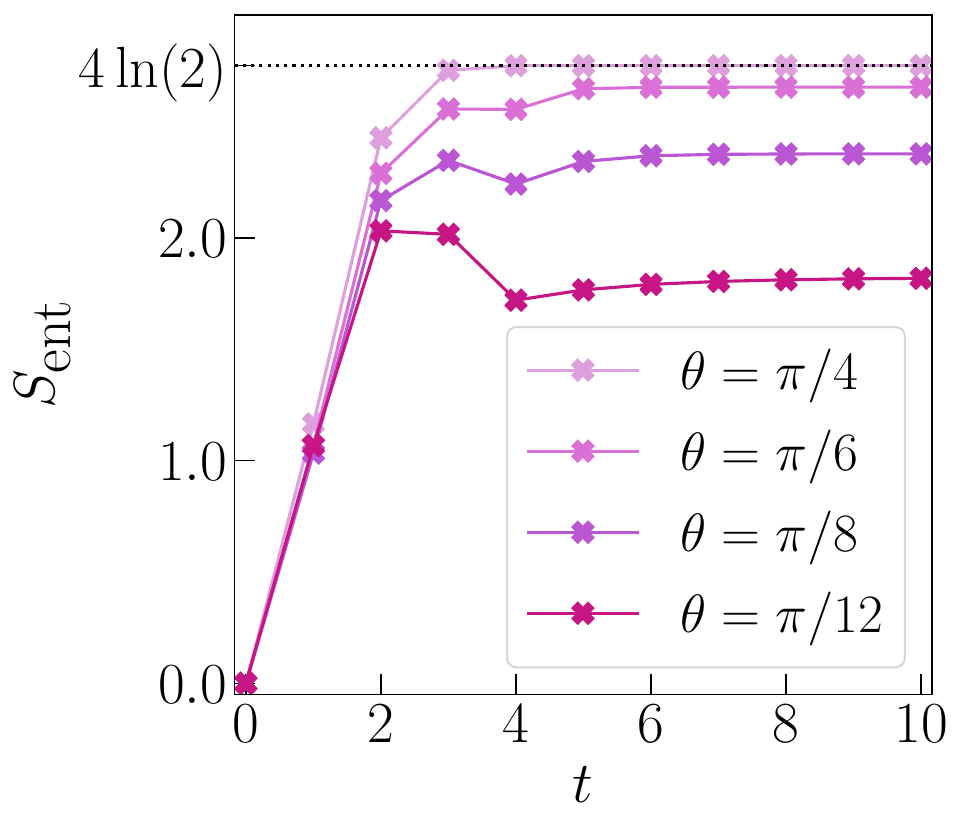} 
\caption{Entanglement dynamics. Left panel: illustration of the joint system, where the right region consists of four sites. Right panel: Growth of subsystem von Neumann entropies for different left initial MPSs, which are characterized by the value of $\theta$. The right initial state and two-site gates are kept the same. The horizontal dotted line marks the maximal entropy $4\ln(2)$ approached by $\theta=\pi/4$.
}
\label{Dynamics}
\end{figure}

As an illustrative example, we report numerical results about the finite-size subsystem entanglement dynamics evolved by Eq. \eqref{eq:q4chi2}. For the left initial state, we choose the following $\chi=2$ MPS, which spans $\mathcal{H}_A$ with $\tilde{q}=2$: $A^{(0)}=
\left( \begin{array}{cc}
\cos(\theta) & \sin(\theta)  \\
0 & 0 
\end{array} \right)$, $A^{(1)}=
\left( \begin{array}{cc}
0 & 0  \\
-\sin(\theta) & \cos(\theta) 
\end{array} \right)$, and $A^{(2,3)}=0$. Here $\theta\in(0,\frac{\pi}{4}]$, which corresponds to the left initial state interpolating between the Greenberger-Horne-Zeilinger state and the cluster state \cite{Raussendorf2001One,Verstraete2004Valence} of basis states $\ket{0}$ and $\ket{1}$. 
As for the right region, we take the initial state as the simple product state $\ket{\Psi_R^j}=\otimes_{x\ge 0}\ket{2}_x$, $j=0,1$, and thus the initial entanglement between regions L and R is zero.
The right subsystem consists of four sites as shown faithfully in the left panel of Fig. \ref{Dynamics}. We construct the two-site gate $U$ by randomly generating unitaries $v,g^{(a)}$, and $f_2^{(a)}$.

Employing the influence matrix method, we can numerically keep track of the full time evolution of the joint system. The joint dynamics is generated by the Floquet operator Eq. \eqref{eq:full_time}, which is constituted by the unitary $\mathbb{U}_R$ along with the boundary quantum channel $\mathcal{M}$ given by Eq.\eqref{eq:Kraus}. 
We compute the von Neumann entanglement entropy between two regions as $S_{\text{ent}}(t)=-\text{Tr}\{\rho_R(t)\ln[\rho_R(t)]\}$.
We emphasize that the scenario here is basically different from that described by Eq. \eqref{eq:Renyi}, where the size of $R$ is infinitely large.
Results for various values of $\theta$ are depicted in the right panel of Fig. \ref{Dynamics}. Following a similar rate of linear growth in the early stage of evolution, the entropies approach the $\theta$-dependent steady values after finite time steps. 
Among the steady values, the maximal entropy $4\ln(2)$ is achieved by the case $\theta=\pi/4$, corresponding to the cluster state as the initial MPS.
Notably, this maximal saturated entropy is still smaller than the maximum entropy allowed by the Hilbert space dimension of the right subsystem $\ln(q^{L_R})=4\ln(4)$, which suggests violating the thermalization toward an infinite-temperature state.
This observation hints at the existence of hidden conservation quantities, which demands further research. 

\textit{Conclusions}---In summary, we have established a systematic approach toward construction of nonintegrable quantum circuits exhibiting exact hidden Markovian subsystem dynamics. 
We introduced new principles beyond dual-unitary circuits allowing for closed-form influence matrices of finite bond dimension, which are formulated into the solvable condition on local unitary gates.
Utilizing the tensor-network method, we demonstrated that the system acts as time-local boundary quantum channels on the subsystem, thus inducing exact hidden Markovian dynamics.
Our constructions have unveiled a novel spacetime duality between the initial state MPS of the system and boundary quantum channels acting on the reduced subsystem.

Our work opens up many avenues for future research, such as the introduction of measurements \cite{Claeys2022Exact} and dissipation \cite{Kos2023Circuits},  and generalizations to different architectures \cite{Prosen2021Many,Claeys2023From} and higher spatial dimensions \cite{Jonay2021Triunitary,Suzuki2022Computational,Milbradt2023Ternary}.
The exact influence matrices also provide valuable analytical tools for studying rich phenomena in quantum many-body dynamics, including quantum chaos \cite{Bertini2018Exact}, scrambling \cite{Bertini2020Scrambling}, and deep thermalization \cite{Ho2022Exact,Claeys2022Exact,Ippoliti2023Dynamical}.

Furthermore, our findings on the exact hidden Markovian property could provide fresh insights into the fundamental understanding of the Markovian approximation.
A valid Markovian approximation often resorts to weak system-bath coupling and the separation of timescales \cite{Davies1974Markovian,Lindblad1976Generators}.
However, in a quantum many-body system, focusing on a subsystem and integrating out the rest usually yields non-Markovian subsystem dynamics,  because the ``bath'' and the subsystem typically undergo the same type of dynamics and thus the corresponding timescales are comparable \cite{Breuer2002Theory}. 
In contrast, in our construction the Markovian property emerges nonperturbatively as a consequence of the solvable condition, even provided the presence of initial genuine quantum correlations between the subsystem and the bath.
Understanding deep relations between these solvable quantum circuits and the Markovian approximation remains an intriguing area for further exploration.

\textit{Acknowledgments}---We thank Tianci Zhou for helpful discussions. This work was supported by the National Natural Science Foundation of China (Grant No. 12125405), and National Key R$\&$D Program of China (No. 2023YFA1406702).

\bibliographystyle{apsrev4-1-title}
\bibliography{Heran_circuit}

\begin{thebibliography}{87}
\bibitem [{zip()}]{zipper_footnote}%
  \BibitemOpen
  \href@noop {} {}\bibinfo {note} {We would like to draw a loose analogy
  between the relationship of zipper conditions and our solvable condition,
  with that of the Bethe ansatz and Yang-Baxter equations. The Bethe ansatz was
  initially proposed to solve exact eigenstates of the 1D Heisenberg spin
  chain, while Yang-Baxter equations capture the crucial point (factorizable
  scattering amplitudes) and the solutions give rise to a wide class of
  solvable models. Simlarly, zipper conditions were used to solve specific
  models, while our solvable condition provides a systematic approach to
  constructing solvable quantum circuits.}\BibitemShut {Stop}%
\end{thebibliography}
\end{document}


\title{Supplementary Materials for: Exact Hidden Markovian Dynamics in Quantum Circuits}

\author{He-Ran Wang}
\affiliation{Institute for Advanced Study, Tsinghua University, Beijing 100084,
People's Republic of China}

\author{Xiao-Yang Yang}
\affiliation{Institute for Advanced Study, Tsinghua University, Beijing 100084,
People's Republic of China}
\affiliation{Department of Physics, Tsinghua University, Beijing 100084, People's
Republic of China}

\author{Zhong Wang}
\email{wangzhongemail@tsinghua.edu.cn}
\affiliation{Institute for Advanced Study, Tsinghua University, Beijing 100084,
People's Republic of China}

\maketitle
\section{Exact influence matrix from the solvable condition}
In this section, we derive the exact influence matrix in the Matrix Product State (MPS) representation from the solvable condition.
We present the proof from two directions: First, we perform straightforward contractions on the 1+1D tensor network to obtain the influence matrix on the time slice (Sec. \ref{sec:contraction}); Second, we show that the obtained influence matrix is indeed the fixed point of the spatial transfer matrix (Sec. \ref{sec:fixed}). 
All the calculations are shown in diagrammatic representations. 

\subsection{Contractions of the tensor network}\label{sec:contraction}
\begin{figure}
\includegraphics[width=1\linewidth]{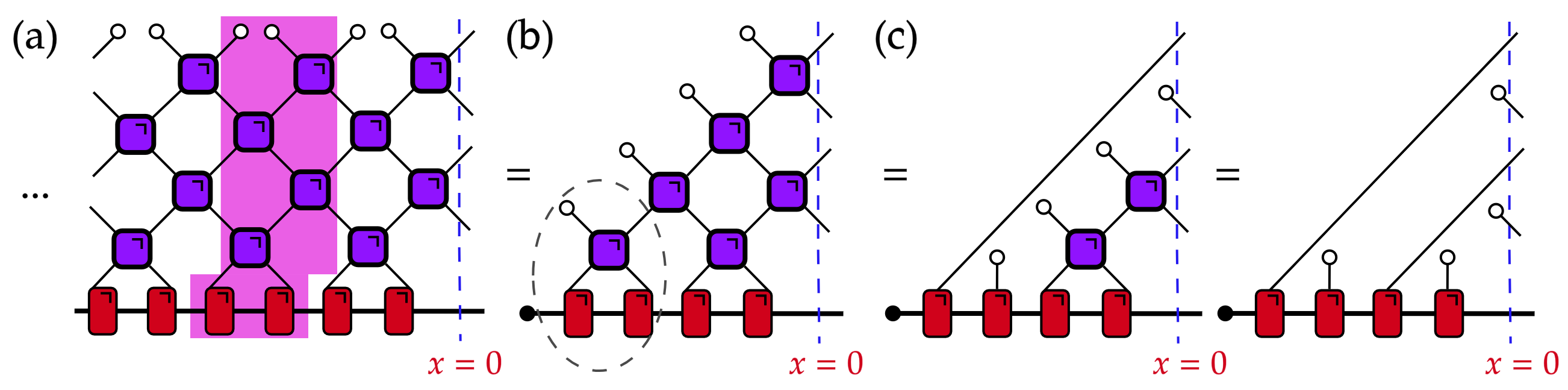}
\caption{Contractions of the tensor network. Total time steps $T=2$. The red shaded region in (a) is a layer of spatial transfer matrix. From (a) to (b), we use two local rules Eq. \eqref{eq:unitary} and Eq. \eqref{eq:canonical} to contract all the tensors within the lightcone. From (b) to (c), we proceed contractions using the solvable condition Eq. \eqref{eq:solvable}.}
\label{Procedure}
\end{figure}

We summarize the procedure of contractions in Fig. \ref{Procedure}. We begin with the tensor network representing the traced time-evolved state in the region L, as depicted in Fig. \ref{Procedure}(a) [see also Fig. 1(b) in the main text]. In the first step, we apply the following two local rules: the unitarity of local gates
\begin{equation}\label{eq:unitary}
\tikzset{every picture/.style={line width=0.75pt}} 
\begin{tikzpicture}[x=0.75pt,y=0.75pt,yscale=-1,xscale=1]
\draw  [fill={rgb, 255:red, 144; green, 19; blue, 254 }  ,fill opacity=1 ][line width=1.5]  (119.2,104.66) .. controls (119.2,102.75) and (120.75,101.2) .. (122.66,101.2) -- (130.74,101.2) .. controls (132.65,101.2) and (134.2,102.75) .. (134.2,104.66) -- (134.2,112.74) .. controls (134.2,114.65) and (132.65,116.2) .. (130.74,116.2) -- (122.66,116.2) .. controls (120.75,116.2) and (119.2,114.65) .. (119.2,112.74) -- cycle ;
\draw  [fill={rgb, 255:red, 144; green, 19; blue, 254 }  ,fill opacity=1 ][line width=0.75]  (126.7,104.7) -- (130.7,104.7) -- (130.7,108.7) ;
\draw [fill={rgb, 255:red, 144; green, 19; blue, 254 }  ,fill opacity=1 ][line width=0.75]    (133.99,115.82) -- (144.2,126.2) ;
\draw [fill={rgb, 255:red, 144; green, 19; blue, 254 }  ,fill opacity=1 ][line width=0.75]    (109.2,91.2) -- (119.7,101.7) ;
\draw [fill={rgb, 255:red, 144; green, 19; blue, 254 }  ,fill opacity=1 ][line width=0.75]    (144.87,91.2) -- (133.98,102.14) ;
\draw [fill={rgb, 255:red, 144; green, 19; blue, 254 }  ,fill opacity=1 ][line width=0.75]    (119.87,114.87) -- (109.2,126.2) ;

\draw  [fill={rgb, 255:red, 255; green, 255; blue, 255 }  ,fill opacity=1 ] (106.4,91.2) .. controls (106.4,89.65) and (107.65,88.4) .. (109.2,88.4) .. controls (110.75,88.4) and (112,89.65) .. (112,91.2) .. controls (112,92.75) and (110.75,94) .. (109.2,94) .. controls (107.65,94) and (106.4,92.75) .. (106.4,91.2) -- cycle ;
\draw    (174.4,94.18) -- (174.2,114.44) -- (174.2,125.4) ;
\draw   (171.6,91.29) .. controls (171.6,89.69) and (172.85,88.4) .. (174.4,88.4) .. controls (175.95,88.4) and (177.2,89.69) .. (177.2,91.29) .. controls (177.2,92.89) and (175.95,94.18) .. (174.4,94.18) .. controls (172.85,94.18) and (171.6,92.89) .. (171.6,91.29) -- cycle ;

\draw  [fill={rgb, 255:red, 255; green, 255; blue, 255 }  ,fill opacity=1 ] (142.07,91.2) .. controls (142.07,89.65) and (143.32,88.4) .. (144.87,88.4) .. controls (146.41,88.4) and (147.67,89.65) .. (147.67,91.2) .. controls (147.67,92.75) and (146.41,94) .. (144.87,94) .. controls (143.32,94) and (142.07,92.75) .. (142.07,91.2) -- cycle ;
\draw    (194.2,94.18) -- (194,114.44) -- (194,125.4) ;
\draw   (191.4,91.29) .. controls (191.4,89.69) and (192.65,88.4) .. (194.2,88.4) .. controls (195.75,88.4) and (197,89.69) .. (197,91.29) .. controls (197,92.89) and (195.75,94.18) .. (194.2,94.18) .. controls (192.65,94.18) and (191.4,92.89) .. (191.4,91.29) -- cycle ;

\draw (149.6,105) node [anchor=north west][inner sep=0.75pt]    {$=$};
\end{tikzpicture},
\end{equation}
and the left-canonical condition of the MPS
\begin{equation}\label{eq:canonical}
\tikzset{every picture/.style={line width=0.75pt}} 
\begin{tikzpicture}[x=0.75pt,y=0.75pt,yscale=-1,xscale=1]

\draw [fill={rgb, 255:red, 144; green, 19; blue, 254 }  ,fill opacity=1 ][line width=0.75]    (90.1,75.1) -- (90,85.67) ;
\draw [line width=1.5]    (74.4,92.5) -- (102.27,92.6) ;
\draw  [fill={rgb, 255:red, 208; green, 2; blue, 27 }  ,fill opacity=1 ] (83.87,83.88) .. controls (83.87,82.62) and (84.89,81.6) .. (86.15,81.6) -- (92.99,81.6) .. controls (94.25,81.6) and (95.27,82.62) .. (95.27,83.88) -- (95.27,98.59) .. controls (95.27,99.85) and (94.25,100.87) .. (92.99,100.87) -- (86.15,100.87) .. controls (84.89,100.87) and (83.87,99.85) .. (83.87,98.59) -- cycle ;
\draw  [line width=0.75]  (88.7,83.83) -- (92.7,83.83) -- (92.7,87.83) ;

\draw  [fill={rgb, 255:red, 0; green, 0; blue, 0 }  ,fill opacity=1 ] (68.8,92.5) .. controls (68.8,90.95) and (70.05,89.7) .. (71.6,89.7) .. controls (73.15,89.7) and (74.4,90.95) .. (74.4,92.5) .. controls (74.4,94.05) and (73.15,95.3) .. (71.6,95.3) .. controls (70.05,95.3) and (68.8,94.05) .. (68.8,92.5) -- cycle ;
\draw [line width=1.5]    (124.2,92.5) -- (152.07,92.6) ;
\draw  [fill={rgb, 255:red, 0; green, 0; blue, 0 }  ,fill opacity=1 ] (121.4,92.5) .. controls (121.4,90.95) and (122.65,89.7) .. (124.2,89.7) .. controls (125.75,89.7) and (127,90.95) .. (127,92.5) .. controls (127,94.05) and (125.75,95.3) .. (124.2,95.3) .. controls (122.65,95.3) and (121.4,94.05) .. (121.4,92.5) -- cycle ;

\draw  [fill={rgb, 255:red, 255; green, 255; blue, 255 }  ,fill opacity=1 ] (87.3,72.3) .. controls (87.3,70.75) and (88.55,69.5) .. (90.1,69.5) .. controls (91.65,69.5) and (92.9,70.75) .. (92.9,72.3) .. controls (92.9,73.85) and (91.65,75.1) .. (90.1,75.1) .. controls (88.55,75.1) and (87.3,73.85) .. (87.3,72.3) -- cycle ;

\draw (104,86.8) node [anchor=north west][inner sep=0.75pt]    {$=$};
\end{tikzpicture}.
\end{equation}
The two local rules allow efficient contractions of the tensor network within the lightcone. More specifically, all the tensors in the space-time region $2t-x>2T$ are reduced to the identities lying along the edge of the lightcone $2t-x=2T$, as shown in Fig. \ref{Procedure}(b). Notice that this step is independent of the solvable condition, and thus is universal for brickwork quantum circuits.

Next, we employ the solvable condition to achieve further simplifications:
\begin{equation}\label{eq:solvable}
\tikzset{every picture/.style={line width=0.75pt}} 
\begin{tikzpicture}[x=0.75pt,y=0.75pt,yscale=-1,xscale=1]
\draw  [fill={rgb, 255:red, 144; green, 19; blue, 254 }  ,fill opacity=1 ][line width=1.5]  (228.33,102.29) .. controls (228.33,100.38) and (229.88,98.83) .. (231.79,98.83) -- (239.87,98.83) .. controls (241.78,98.83) and (243.33,100.38) .. (243.33,102.29) -- (243.33,110.37) .. controls (243.33,112.28) and (241.78,113.83) .. (239.87,113.83) -- (231.79,113.83) .. controls (229.88,113.83) and (228.33,112.28) .. (228.33,110.37) -- cycle ;
\draw  [fill={rgb, 255:red, 144; green, 19; blue, 254 }  ,fill opacity=1 ][line width=0.75]  (235.83,102.33) -- (239.83,102.33) -- (239.83,106.33) ;
\draw [fill={rgb, 255:red, 144; green, 19; blue, 254 }  ,fill opacity=1 ][line width=0.75]    (243.13,113.45) -- (253.33,123.83) ;
\draw [fill={rgb, 255:red, 144; green, 19; blue, 254 }  ,fill opacity=1 ][line width=0.75]    (218.33,88.83) -- (228.83,99.33) ;
\draw [fill={rgb, 255:red, 144; green, 19; blue, 254 }  ,fill opacity=1 ][line width=0.75]    (254,88.83) -- (243.11,99.78) ;
\draw [fill={rgb, 255:red, 144; green, 19; blue, 254 }  ,fill opacity=1 ][line width=0.75]    (229,112.5) -- (218.33,123.83) ;

\draw  [fill={rgb, 255:red, 255; green, 255; blue, 255 }  ,fill opacity=1 ] (215.53,88.83) .. controls (215.53,87.29) and (216.79,86.03) .. (218.33,86.03) .. controls (219.88,86.03) and (221.13,87.29) .. (221.13,88.83) .. controls (221.13,90.38) and (219.88,91.63) .. (218.33,91.63) .. controls (216.79,91.63) and (215.53,90.38) .. (215.53,88.83) -- cycle ;
\draw [line width=1.5]    (208.5,81.94) -- (208.57,140.12) ;
\draw  [fill={rgb, 255:red, 208; green, 2; blue, 27 }  ,fill opacity=1 ] (215.89,120.33) .. controls (217.15,120.33) and (218.17,121.35) .. (218.17,122.61) -- (218.17,129.45) .. controls (218.17,130.71) and (217.15,131.73) .. (215.89,131.73) -- (201.18,131.73) .. controls (199.92,131.73) and (198.9,130.71) .. (198.9,129.45) -- (198.9,122.61) .. controls (198.9,121.35) and (199.92,120.33) .. (201.18,120.33) -- cycle ;
\draw  [line width=0.75]  (215.94,125.17) -- (215.94,129.17) -- (211.94,129.17) ;

\draw [line width=1.5]    (356.5,81.94) -- (356.57,140.12) ;
\draw  [fill={rgb, 255:red, 208; green, 2; blue, 27 }  ,fill opacity=1 ] (363.89,91.33) .. controls (365.15,91.33) and (366.17,92.35) .. (366.17,93.61) -- (366.17,100.45) .. controls (366.17,101.71) and (365.15,102.73) .. (363.89,102.73) -- (349.18,102.73) .. controls (347.92,102.73) and (346.9,101.71) .. (346.9,100.45) -- (346.9,93.61) .. controls (346.9,92.35) and (347.92,91.33) .. (349.18,91.33) -- cycle ;
\draw  [line width=0.75]  (363.94,96.17) -- (363.94,100.17) -- (359.94,100.17) ;

\draw    (366.5,96.13) -- (388,96.12) ;
\draw [fill={rgb, 255:red, 255; green, 255; blue, 255 }  ,fill opacity=1 ]   (366.33,123.83) -- (388.83,123.82) ;
\draw  [fill={rgb, 255:red, 255; green, 255; blue, 255 }  ,fill opacity=1 ] (363.53,123.83) .. controls (363.53,122.29) and (364.79,121.03) .. (366.33,121.03) .. controls (367.88,121.03) and (369.13,122.29) .. (369.13,123.83) .. controls (369.13,125.38) and (367.88,126.63) .. (366.33,126.63) .. controls (364.79,126.63) and (363.53,125.38) .. (363.53,123.83) -- cycle ;
\draw [fill={rgb, 255:red, 144; green, 19; blue, 254 }  ,fill opacity=1 ][line width=0.75]    (310.6,108.37) -- (325.07,123.47) ;
\draw [fill={rgb, 255:red, 144; green, 19; blue, 254 }  ,fill opacity=1 ][line width=0.75]    (293.33,88.86) -- (307.47,104.27) ;
\draw [fill={rgb, 255:red, 144; green, 19; blue, 254 }  ,fill opacity=1 ][line width=0.75]    (324.67,89.07) -- (293.33,123.86) ;
\draw  [fill={rgb, 255:red, 255; green, 255; blue, 255 }  ,fill opacity=1 ] (290.53,88.86) .. controls (290.53,87.31) and (291.79,86.06) .. (293.33,86.06) .. controls (294.88,86.06) and (296.13,87.31) .. (296.13,88.86) .. controls (296.13,90.41) and (294.88,91.66) .. (293.33,91.66) .. controls (291.79,91.66) and (290.53,90.41) .. (290.53,88.86) -- cycle ;
\draw [line width=1.5]    (283.5,81.97) -- (283.57,140.15) ;
\draw  [fill={rgb, 255:red, 208; green, 2; blue, 27 }  ,fill opacity=1 ] (290.89,120.36) .. controls (292.15,120.36) and (293.17,121.38) .. (293.17,122.64) -- (293.17,129.48) .. controls (293.17,130.74) and (292.15,131.76) .. (290.89,131.76) -- (276.18,131.76) .. controls (274.92,131.76) and (273.9,130.74) .. (273.9,129.48) -- (273.9,122.64) .. controls (273.9,121.38) and (274.92,120.36) .. (276.18,120.36) -- cycle ;
\draw  [line width=0.75]  (290.94,125.19) -- (290.94,129.19) -- (286.94,129.19) ;

\draw (253.8,103.14) node [anchor=north west][inner sep=0.75pt]    {$=$};
\draw (323,102.74) node [anchor=north west][inner sep=0.75pt]    {$=$};
\end{tikzpicture}.
\end{equation}
For instance, we can apply this condition on the tensor located at the lower-left corner in Fig. \ref{Procedure}(b) (circled by the grey dotted line), which gives
\begin{equation}
\tikzset{every picture/.style={line width=0.75pt}} 
\begin{tikzpicture}[x=0.75pt,y=0.75pt,yscale=-1,xscale=1]
\draw  [fill={rgb, 255:red, 144; green, 19; blue, 254 }  ,fill opacity=1 ][line width=1.5]  (172.33,78.69) .. controls (172.33,76.78) and (173.88,75.23) .. (175.79,75.23) -- (183.87,75.23) .. controls (185.78,75.23) and (187.33,76.78) .. (187.33,78.69) -- (187.33,86.77) .. controls (187.33,88.68) and (185.78,90.23) .. (183.87,90.23) -- (175.79,90.23) .. controls (173.88,90.23) and (172.33,88.68) .. (172.33,86.77) -- cycle ;
\draw  [fill={rgb, 255:red, 144; green, 19; blue, 254 }  ,fill opacity=1 ][line width=0.75]  (179.83,78.73) -- (183.83,78.73) -- (183.83,82.73) ;
\draw [fill={rgb, 255:red, 144; green, 19; blue, 254 }  ,fill opacity=1 ][line width=0.75]    (187.13,89.85) -- (197.33,100.23) ;
\draw [fill={rgb, 255:red, 144; green, 19; blue, 254 }  ,fill opacity=1 ][line width=0.75]    (162.33,65.23) -- (172.83,75.73) ;
\draw [fill={rgb, 255:red, 144; green, 19; blue, 254 }  ,fill opacity=1 ][line width=0.75]    (198,65.23) -- (187.11,76.18) ;
\draw [fill={rgb, 255:red, 144; green, 19; blue, 254 }  ,fill opacity=1 ][line width=0.75]    (173,88.9) -- (162.33,100.23) ;
\draw [line width=1.5]    (145.94,108.97) -- (211.13,109.1) ;
\draw  [fill={rgb, 255:red, 208; green, 2; blue, 27 }  ,fill opacity=1 ] (161.47,101.68) .. controls (161.47,100.42) and (162.49,99.4) .. (163.75,99.4) -- (170.59,99.4) .. controls (171.85,99.4) and (172.87,100.42) .. (172.87,101.68) -- (172.87,116.39) .. controls (172.87,117.65) and (171.85,118.67) .. (170.59,118.67) -- (163.75,118.67) .. controls (162.49,118.67) and (161.47,117.65) .. (161.47,116.39) -- cycle ;
\draw  [line width=0.75]  (166.3,101.63) -- (170.3,101.63) -- (170.3,105.63) ;

\draw  [fill={rgb, 255:red, 208; green, 2; blue, 27 }  ,fill opacity=1 ] (186.33,101.68) .. controls (186.33,100.42) and (187.35,99.4) .. (188.61,99.4) -- (195.45,99.4) .. controls (196.71,99.4) and (197.73,100.42) .. (197.73,101.68) -- (197.73,116.39) .. controls (197.73,117.65) and (196.71,118.67) .. (195.45,118.67) -- (188.61,118.67) .. controls (187.35,118.67) and (186.33,117.65) .. (186.33,116.39) -- cycle ;
\draw  [line width=0.75]  (191.17,101.63) -- (195.17,101.63) -- (195.17,105.63) ;

\draw  [fill={rgb, 255:red, 255; green, 255; blue, 255 }  ,fill opacity=1 ] (159.53,65.23) .. controls (159.53,63.69) and (160.79,62.43) .. (162.33,62.43) .. controls (163.88,62.43) and (165.13,63.69) .. (165.13,65.23) .. controls (165.13,66.78) and (163.88,68.03) .. (162.33,68.03) .. controls (160.79,68.03) and (159.53,66.78) .. (159.53,65.23) -- cycle ;
\draw [fill={rgb, 255:red, 144; green, 19; blue, 254 }  ,fill opacity=1 ][line width=0.75]    (277.83,89.03) -- (277.87,99.53) ;
\draw [fill={rgb, 255:red, 144; green, 19; blue, 254 }  ,fill opacity=1 ][line width=0.75]    (286.17,65.93) -- (252.33,99.23) ;
\draw [line width=1.5]    (231.94,108.97) -- (297.13,109.1) ;
\draw  [fill={rgb, 255:red, 208; green, 2; blue, 27 }  ,fill opacity=1 ] (247.47,101.68) .. controls (247.47,100.42) and (248.49,99.4) .. (249.75,99.4) -- (256.59,99.4) .. controls (257.85,99.4) and (258.87,100.42) .. (258.87,101.68) -- (258.87,116.39) .. controls (258.87,117.65) and (257.85,118.67) .. (256.59,118.67) -- (249.75,118.67) .. controls (248.49,118.67) and (247.47,117.65) .. (247.47,116.39) -- cycle ;
\draw  [line width=0.75]  (252.3,101.63) -- (256.3,101.63) -- (256.3,105.63) ;

\draw  [fill={rgb, 255:red, 208; green, 2; blue, 27 }  ,fill opacity=1 ] (272.33,101.68) .. controls (272.33,100.42) and (273.35,99.4) .. (274.61,99.4) -- (281.45,99.4) .. controls (282.71,99.4) and (283.73,100.42) .. (283.73,101.68) -- (283.73,116.39) .. controls (283.73,117.65) and (282.71,118.67) .. (281.45,118.67) -- (274.61,118.67) .. controls (273.35,118.67) and (272.33,117.65) .. (272.33,116.39) -- cycle ;
\draw  [line width=0.75]  (277.17,101.63) -- (281.17,101.63) -- (281.17,105.63) ;

\draw  [fill={rgb, 255:red, 255; green, 255; blue, 255 }  ,fill opacity=1 ] (275.03,86.23) .. controls (275.03,84.69) and (276.29,83.43) .. (277.83,83.43) .. controls (279.38,83.43) and (280.63,84.69) .. (280.63,86.23) .. controls (280.63,87.78) and (279.38,89.03) .. (277.83,89.03) .. controls (276.29,89.03) and (275.03,87.78) .. (275.03,86.23) -- cycle ;

\draw (215,86) node [anchor=north west][inner sep=0.75pt]    {$=$};
\end{tikzpicture}.
\end{equation}
We perform such contractions along the edge of the lightcone $2t-x=2T$ to shift the identities to the next edge, as shown in Fig. \ref{Procedure} (c).
By repeating this ``lightcone decimation'' procedure for $T$ times, we arrive at the closed-form influence matrix with outer legs across the time slice, the same as Fig. 1(c) in the main text.

The contractions imply a novel manifestation of \textit{space-time duality} between the initial state MPS and the influence matrix: The local MPS located at an even site $x$ in the initial state exactly corresponds to the local MPS in the influence matrix at $t=-x/2$. 

\subsection{Fixed point condition}\label{sec:fixed}
In quantum circuits respecting shift invariance in the space direction, the influence matrix of a global system can be identified as the fixed point of the spatial transfer matrix \cite{Banuls2009Matrix,Hermes2012Tensor,Hastings2015Connecting,Lerose2021Influence}. In our brickwork architecture which is two-site shift invariant, the spatial transfer matrix [visualized in red shaded region in Fig. \ref{Procedure}(a)] involves two layers.
To verify the fixed point condition, we contract the spatial transfer matrix with the obtained influence matrix from the right hand side, which yields
\begin{equation}
\tikzset{every picture/.style={line width=0.75pt}} 

\end{equation}
In the first two equalities, we only exploit the solvable condition Eq. \eqref{eq:solvable}, while the last equality is based on the left-canonical form [Eq. \eqref{eq:canonical}].
We have proved that the MPS influence matrix is the fixed point of the spatial transfer matrix.

\subsection{Two-site shift-invariant initial MPS}\label{sec:twosite}
Here we discuss a class of two-site shift-invariant initial MPS which also leads to exact influence matrices under the solvable condition Eq. \eqref{eq:solvable}. We consider the following left initial state in the alternating MPS form:
\begin{equation}
\tikzset{every picture/.style={line width=0.75pt}} 
\begin{tikzpicture}[x=0.75pt,y=0.75pt,yscale=-1,xscale=1]

\draw [fill={rgb, 255:red, 144; green, 19; blue, 254 }  ,fill opacity=1 ][line width=0.75]    (159.9,71.9) -- (159.95,80.8) ;
\draw [line width=1.5]    (230.98,91.64) -- (242.98,91.47) ;
\draw [line width=1.5]    (148,91.7) -- (249.57,91.7) ;
\draw  [fill={rgb, 255:red, 208; green, 2; blue, 27 }  ,fill opacity=1 ] (154.47,83.08) .. controls (154.47,81.82) and (155.49,80.8) .. (156.75,80.8) -- (163.59,80.8) .. controls (164.85,80.8) and (165.87,81.82) .. (165.87,83.08) -- (165.87,97.79) .. controls (165.87,99.05) and (164.85,100.07) .. (163.59,100.07) -- (156.75,100.07) .. controls (155.49,100.07) and (154.47,99.05) .. (154.47,97.79) -- cycle ;
\draw  [fill={rgb, 255:red, 208; green, 2; blue, 27 }  ,fill opacity=1 ][line width=0.75]  (159.3,83.03) -- (163.3,83.03) -- (163.3,87.03) ;

\draw  [fill={rgb, 255:red, 2; green, 100; blue, 200 }  ,fill opacity=1 ] (228.93,83.08) .. controls (228.93,81.82) and (229.95,80.8) .. (231.21,80.8) -- (238.05,80.8) .. controls (239.31,80.8) and (240.33,81.82) .. (240.33,83.08) -- (240.33,97.79) .. controls (240.33,99.05) and (239.31,100.07) .. (238.05,100.07) -- (231.21,100.07) .. controls (229.95,100.07) and (228.93,99.05) .. (228.93,97.79) -- cycle ;
\draw  [fill={rgb, 255:red, 2; green, 100; blue, 200 }  ,fill opacity=1 ][line width=0.75]  (233.77,83.03) -- (237.77,83.03) -- (237.77,87.03) ;

\draw  [fill={rgb, 255:red, 208; green, 2; blue, 27 }  ,fill opacity=1 ] (203.33,83.08) .. controls (203.33,81.82) and (204.35,80.8) .. (205.61,80.8) -- (212.45,80.8) .. controls (213.71,80.8) and (214.73,81.82) .. (214.73,83.08) -- (214.73,97.79) .. controls (214.73,99.05) and (213.71,100.07) .. (212.45,100.07) -- (205.61,100.07) .. controls (204.35,100.07) and (203.33,99.05) .. (203.33,97.79) -- cycle ;
\draw  [fill={rgb, 255:red, 208; green, 2; blue, 27 }  ,fill opacity=1 ][line width=0.75]  (208.17,83.03) -- (212.17,83.03) -- (212.17,87.03) ;

\draw  [fill={rgb, 255:red, 2; green, 100; blue, 200 }  ,fill opacity=1 ] (179.33,83.08) .. controls (179.33,81.82) and (180.35,80.8) .. (181.61,80.8) -- (188.45,80.8) .. controls (189.71,80.8) and (190.73,81.82) .. (190.73,83.08) -- (190.73,97.79) .. controls (190.73,99.05) and (189.71,100.07) .. (188.45,100.07) -- (181.61,100.07) .. controls (180.35,100.07) and (179.33,99.05) .. (179.33,97.79) -- cycle ;
\draw  [fill={rgb, 255:red, 2; green, 100; blue, 200 }  ,fill opacity=1 ][line width=0.75]  (184.17,83.03) -- (188.17,83.03) -- (188.17,87.03) ;

\draw [fill={rgb, 255:red, 144; green, 19; blue, 254 }  ,fill opacity=1 ][line width=0.75]    (234.3,71.9) -- (234.35,80.8) ;
\draw [fill={rgb, 255:red, 144; green, 19; blue, 254 }  ,fill opacity=1 ][line width=0.75]    (208.7,71.5) -- (208.75,80.4) ;
\draw [fill={rgb, 255:red, 144; green, 19; blue, 254 }  ,fill opacity=1 ][line width=0.75]    (185.1,71.9) -- (185.15,80.8) ;

\draw (140.97,93) node   [align=left] {\begin{minipage}[lt]{14.55pt}\setlength\topsep{0pt}
...
\end{minipage}};
\draw (252.47,87.87) node [anchor=north west][inner sep=0.75pt]  [font=\scriptsize]  {$j,j'$,};
\end{tikzpicture}
\end{equation}
where the new tensor in blue is defined as
\begin{equation}
\tikzset{every picture/.style={line width=0.75pt}} 
\begin{tikzpicture}[x=0.75pt,y=0.75pt,yscale=-1,xscale=1]

\draw [line width=1.5]    (120,143.04) -- (149,143.04) ;
\draw  [fill={rgb, 255:red, 2; green, 100; blue, 200 }  ,fill opacity=1 ] (128.47,135.68) .. controls (128.47,134.42) and (129.49,133.4) .. (130.75,133.4) -- (137.59,133.4) .. controls (138.85,133.4) and (139.87,134.42) .. (139.87,135.68) -- (139.87,150.39) .. controls (139.87,151.65) and (138.85,152.67) .. (137.59,152.67) -- (130.75,152.67) .. controls (129.49,152.67) and (128.47,151.65) .. (128.47,150.39) -- cycle ;
\draw  [fill={rgb, 255:red, 2; green, 100; blue, 200 }  ,fill opacity=1 ][line width=0.75]  (133.3,135.63) -- (137.3,135.63) -- (137.3,139.63) ;

\draw [fill={rgb, 255:red, 144; green, 19; blue, 254 }  ,fill opacity=1 ][line width=0.75]    (134.2,124) -- (134.25,132.9) ;

\draw (124.8,112.21) node [anchor=north west][inner sep=0.75pt]  [font=\scriptsize]  {$a,a'$};
\draw (101.4,135.61) node [anchor=north west][inner sep=0.75pt]  [font=\scriptsize]  {$j,j'$};
\draw (150.4,136.61) node [anchor=north west][inner sep=0.75pt]  [font=\scriptsize]  {$k,k'$};
\draw (175.57,124.7) node [anchor=north west][inner sep=0.75pt]    {$=B_{jk}^{( a)} B{_{j'k'}^{( a')}}^{*} .$};
\end{tikzpicture}
\end{equation}
In such an initial state, the even (odd) site is associated with the tensor $A$($B$) (remind that we have set the leftmost site of the right region to $x=0$). In this case, the matrices $A^{(a)}$ and $B^{(b)}$ are not necessarily square matrices; they can be $\chi\times\chi'$ and $\chi'\times\chi$, respectively. We still require that the MPS is injective and in the left-canonical form, which reads
\begin{equation}
    \sum_{a,b=0}^{q-1}{(A^{(a)}B^{(b)})}^\dagger A^{(a)}B^{(b)}=I_\chi.
\end{equation}

Notice that the solvable condition Eq. \eqref{eq:solvable} we imposed is irrelevant of the tensor $B$. Following the same approach presented in Fig. \ref{Procedure}, one can show that tracing out the left region results in the following influence matrix (an example for $t=2$):
\begin{equation}
\tikzset{every picture/.style={line width=0.75pt}} 
\begin{tikzpicture}[x=0.75pt,y=0.75pt,yscale=-1,xscale=1]

\draw [fill={rgb, 255:red, 144; green, 19; blue, 254 }  ,fill opacity=1 ][line width=0.75]    (344.53,30.26) -- (304.67,30.67) ;
\draw [fill={rgb, 255:red, 144; green, 19; blue, 254 }  ,fill opacity=1 ][line width=0.75]    (327.44,41.93) -- (344.25,41.92) ;
\draw [fill={rgb, 255:red, 144; green, 19; blue, 254 }  ,fill opacity=1 ][line width=0.75]    (343.25,90.32) -- (326.47,90.05) ;
\draw  [fill={rgb, 255:red, 255; green, 255; blue, 255 }  ,fill opacity=1 ] (321.84,41.93) .. controls (321.84,40.39) and (323.1,39.13) .. (324.64,39.13) .. controls (326.19,39.13) and (327.44,40.39) .. (327.44,41.93) .. controls (327.44,43.48) and (326.19,44.73) .. (324.64,44.73) .. controls (323.1,44.73) and (321.84,43.48) .. (321.84,41.93) -- cycle ;
\draw  [fill={rgb, 255:red, 0; green, 0; blue, 0 }  ,fill opacity=1 ] (294.98,16.91) .. controls (294.98,15.37) and (296.23,14.11) .. (297.78,14.11) .. controls (299.32,14.11) and (300.58,15.37) .. (300.58,16.91) .. controls (300.58,18.46) and (299.32,19.71) .. (297.78,19.71) .. controls (296.23,19.71) and (294.98,18.46) .. (294.98,16.91) -- cycle ;
\draw [line width=1.5]    (297.47,82.99) -- (297.44,135.27) -- (342.85,135.05) ;
\draw  [fill={rgb, 255:red, 255; green, 255; blue, 255 }  ,fill opacity=1 ] (321.87,90.05) .. controls (321.87,88.5) and (323.12,87.25) .. (324.67,87.25) .. controls (326.21,87.25) and (327.47,88.5) .. (327.47,90.05) .. controls (327.47,91.59) and (326.21,92.85) .. (324.67,92.85) .. controls (323.12,92.85) and (321.87,91.59) .. (321.87,90.05) -- cycle ;
\draw [fill={rgb, 255:red, 144; green, 19; blue, 254 }  ,fill opacity=1 ][line width=0.75]    (343.83,77.43) -- (303.17,77.19) ;
\draw [line width=1.5]    (297.56,16.57) -- (297.47,82.99) ;
\draw  [fill={rgb, 255:red, 208; green, 2; blue, 27 }  ,fill opacity=1 ] (305.72,71.66) .. controls (306.98,71.66) and (308,72.68) .. (308,73.94) -- (308,80.78) .. controls (308,82.04) and (306.98,83.06) .. (305.72,83.06) -- (291.01,83.06) .. controls (289.75,83.06) and (288.73,82.04) .. (288.73,80.78) -- (288.73,73.94) .. controls (288.73,72.68) and (289.75,71.66) .. (291.01,71.66) -- cycle ;
\draw  [line width=0.75]  (305.77,76.49) -- (305.77,80.49) -- (301.77,80.49) ;

\draw  [fill={rgb, 255:red, 208; green, 2; blue, 27 }  ,fill opacity=1 ] (305.72,25.66) .. controls (306.98,25.66) and (308,26.68) .. (308,27.94) -- (308,34.78) .. controls (308,36.04) and (306.98,37.06) .. (305.72,37.06) -- (291.01,37.06) .. controls (289.75,37.06) and (288.73,36.04) .. (288.73,34.78) -- (288.73,27.94) .. controls (288.73,26.68) and (289.75,25.66) .. (291.01,25.66) -- cycle ;
\draw  [line width=0.75]  (305.77,30.49) -- (305.77,34.49) -- (301.77,34.49) ;

\draw  [fill={rgb, 255:red, 2; green, 100; blue, 200 }  ,fill opacity=1 ] (305.72,48.26) .. controls (306.98,48.26) and (308,49.28) .. (308,50.54) -- (308,57.38) .. controls (308,58.64) and (306.98,59.66) .. (305.72,59.66) -- (291.01,59.66) .. controls (289.75,59.66) and (288.73,58.64) .. (288.73,57.38) -- (288.73,50.54) .. controls (288.73,49.28) and (289.75,48.26) .. (291.01,48.26) -- cycle ;
\draw  [fill={rgb, 255:red, 2; green, 100; blue, 200 }  ,fill opacity=1 ][line width=0.75]  (305.77,53.09) -- (305.77,57.09) -- (301.77,57.09) ;

\draw  [fill={rgb, 255:red, 2; green, 100; blue, 200 }  ,fill opacity=1 ] (305.72,95.66) .. controls (306.98,95.66) and (308,96.68) .. (308,97.94) -- (308,104.78) .. controls (308,106.04) and (306.98,107.06) .. (305.72,107.06) -- (291.01,107.06) .. controls (289.75,107.06) and (288.73,106.04) .. (288.73,104.78) -- (288.73,97.94) .. controls (288.73,96.68) and (289.75,95.66) .. (291.01,95.66) -- cycle ;
\draw  [fill={rgb, 255:red, 2; green, 100; blue, 200 }  ,fill opacity=1 ][line width=0.75]  (305.77,100.49) -- (305.77,104.49) -- (301.77,104.49) ;

\draw [fill={rgb, 255:red, 144; green, 19; blue, 254 }  ,fill opacity=1 ][line width=0.75]    (307.84,53.93) -- (318.54,54.16) ;
\draw [fill={rgb, 255:red, 144; green, 19; blue, 254 }  ,fill opacity=1 ][line width=0.75]    (308.24,102.33) -- (318.94,102.56) ;
\draw  [fill={rgb, 255:red, 255; green, 255; blue, 255 }  ,fill opacity=1 ] (315.74,54.16) .. controls (315.74,52.61) and (317,51.36) .. (318.54,51.36) .. controls (320.09,51.36) and (321.34,52.61) .. (321.34,54.16) .. controls (321.34,55.7) and (320.09,56.96) .. (318.54,56.96) .. controls (317,56.96) and (315.74,55.7) .. (315.74,54.16) -- cycle ;
\draw  [fill={rgb, 255:red, 255; green, 255; blue, 255 }  ,fill opacity=1 ] (316.14,102.56) .. controls (316.14,101.01) and (317.4,99.76) .. (318.94,99.76) .. controls (320.49,99.76) and (321.74,101.01) .. (321.74,102.56) .. controls (321.74,104.1) and (320.49,105.36) .. (318.94,105.36) .. controls (317.4,105.36) and (316.14,104.1) .. (316.14,102.56) -- cycle ;
\end{tikzpicture}
\end{equation}

\section{Boundary quantum channel}
\subsection{Kraus-operator representation}
In this subection, we derive the concrete expressions of the time-local boundary quantum channel defined by
\begin{equation}
\tikzset{every picture/.style={line width=0.75pt}} 
\begin{tikzpicture}[x=0.75pt,y=0.75pt,yscale=-1,xscale=1]
\draw    (274.57,90.18) -- (283.07,90.22) ;
\draw [line width=1.5]    (248.82,51.39) -- (248.74,109.57) ;
\draw  [fill={rgb, 255:red, 208; green, 2; blue, 27 }  ,fill opacity=1 ] (255.94,60.77) .. controls (257.2,60.76) and (258.22,61.78) .. (258.22,63.04) -- (258.24,69.88) .. controls (258.24,71.14) and (257.22,72.16) .. (255.97,72.17) -- (241.26,72.2) .. controls (240,72.21) and (238.98,71.19) .. (238.97,69.93) -- (238.96,63.09) .. controls (238.95,61.83) and (239.97,60.81) .. (241.23,60.8) -- cycle ;
\draw  [line width=0.75]  (256,65.6) -- (256.01,69.6) -- (252.01,69.61) ;

\draw    (258.56,66.56) -- (283.06,66.61) ;
\draw [fill={rgb, 255:red, 255; green, 255; blue, 255 }  ,fill opacity=1 ]   (258.47,90.26) -- (264.97,90.18) ;
\draw  [fill={rgb, 255:red, 255; green, 255; blue, 255 }  ,fill opacity=1 ] (261.97,90.18) .. controls (261.96,88.63) and (263.21,87.38) .. (264.76,87.37) .. controls (266.31,87.37) and (267.56,88.62) .. (267.57,90.16) .. controls (267.57,91.71) and (266.32,92.97) .. (264.77,92.97) .. controls (263.23,92.98) and (261.97,91.72) .. (261.97,90.18) -- cycle ;
\draw  [fill={rgb, 255:red, 208; green, 2; blue, 27 }  ,fill opacity=1 ] (255.72,84.37) .. controls (256.98,84.36) and (258,85.38) .. (258,86.64) -- (258.02,93.48) .. controls (258.02,94.74) and (257,95.76) .. (255.74,95.77) -- (241.04,95.8) .. controls (239.78,95.81) and (238.76,94.79) .. (238.75,93.53) -- (238.74,86.69) .. controls (238.73,85.43) and (239.75,84.41) .. (241.01,84.4) -- cycle ;
\draw  [line width=0.75]  (255.78,89.2) -- (255.79,93.2) -- (251.79,93.21) ;

\draw  [fill={rgb, 255:red, 255; green, 255; blue, 255 }  ,fill opacity=1 ] (270.77,90.19) .. controls (270.76,88.64) and (272.01,87.38) .. (273.56,87.38) .. controls (275.11,87.38) and (276.36,88.63) .. (276.37,90.17) .. controls (276.37,91.72) and (275.12,92.98) .. (273.57,92.98) .. controls (272.03,92.98) and (270.77,91.73) .. (270.77,90.19) -- cycle ;
\draw  [color={rgb, 255:red, 0; green, 0; blue, 0 }  ,draw opacity=1 ][fill={rgb, 255:red, 208; green, 2; blue, 27 }  ,fill opacity=1 ][line width=1.5]  (183.2,79.42) .. controls (183.2,74.58) and (187.12,70.67) .. (191.95,70.67) .. controls (196.78,70.67) and (200.7,74.58) .. (200.7,79.42) .. controls (200.7,84.25) and (196.78,88.17) .. (191.95,88.17) .. controls (187.12,88.17) and (183.2,84.25) .. (183.2,79.42) -- cycle ;
\draw [fill={rgb, 255:red, 144; green, 19; blue, 254 }  ,fill opacity=1 ][line width=0.75]    (197.67,85.42) -- (208.48,95.53) ;
\draw [fill={rgb, 255:red, 144; green, 19; blue, 254 }  ,fill opacity=1 ][line width=0.75]    (210.64,61.57) -- (198.67,72.42) ;
\draw [line width=1.5]    (185.48,85.98) -- (170.05,103.98) ;
\draw [line width=1.5]    (168.12,57.55) -- (186.05,74.24) ;
\draw  [line width=0.75]  (191.9,75.37) -- (195.9,75.37) -- (195.9,79.37) ;

\draw (215.06,70.01) node [anchor=north west][inner sep=0.75pt]    {$=$};
\end{tikzpicture}.
\end{equation}
The action of quantum channel can be directly read out from the diagram following the down-to-up direction:
\begin{equation}\label{eq:channel}
    \mathcal{M}(\tilde{\rho}_R)=\sum_{b,b'=0}^{q-1}A^{(b)}(\sum_{a=0}^{q-1}A^{(a)}\text{Tr}_{x=0}[\tilde{\rho}_R]{A^{(a)}}^\dagger){A^{(b')}}^\dagger\bigotimes (\ket{b}\bra{b'})_{x=0}.
\end{equation}
We shall bring this expression into the standard Kraus representation. To this end, we rewrite the trace operation over the Hilbert space at $x=0$ as $\text{Tr}_{x=0}[\tilde{\rho}_R]=\sum_{a'=0}^{q-1}\bra{a'}\tilde{\rho}_R\ket{a'}$. Plugging this decomposition into Eq. \eqref{eq:channel} leads to 
\begin{equation}
    \mathcal{M}(\tilde{\rho}_R)=\sum_{a,a',b,b'=0}^{q-1}[A^{(b)}A^{(a)}\bigotimes (\ket{b}\bra{a'})_{x=0}]\tilde{\rho}_R[{A^{(a)}}^\dagger{A^{(b')}}^\dagger\bigotimes (\ket{a'}\bra{b'})_{x=0}].
\end{equation}
The summation is over $a,a',b,b'$. However, notice that the index $b$ and $b'$ appear in different sides of $\tilde{\rho}_R$ independently. Therefore, we can recast the operators on the single side of $\tilde{\rho}_R$ into the Kraus operator by summing over $b$:
\begin{equation}
    K_{a,a'}=\sum_{b=0}^{q-1} A^{(b)}A^{(a)}\bigotimes(\ket{b}\bra{a'})_{x=0}.
\end{equation}
Consequently, Eq. \eqref{eq:channel} becomes $\mathcal{M}(\tilde{\rho}_R)=\sum_{a,a'=0}^{q-1} K_{a,a'}\tilde{\rho}_RK_{a,a'}^\dagger$. 
We can check the trace-preserving property of Kraus operators as follows:
\begin{align}
    \sum_{a,a'=0}^{q-1}K_{a,a'}^\dagger K_{a,a'}=&\sum_{a,a',b,b'=0}^{q-1}{A^{(a)}}^\dagger{A^{(b)}}^\dagger A^{(b')}A^{(a)}\bigotimes \ket{a'}\bra{b}b'\rangle\bra{a'}\nonumber\\
    =&(\sum_{a,b,b'=0}^{q-1}\delta_{b,b'}{A^{(a)}}^\dagger{A^{(b)}}^\dagger A^{(b')}A^{(a)})\bigotimes (\sum_{a'=0}^{q-1}\ket{a'}\bra{a'})\nonumber\\
    =&\sum_{a=0}^{q-1}{A^{(a)}}^\dagger(\sum_{b=0}^{q-1}{A^{(b)}}^\dagger A^{(b)})A^{(a)}\bigotimes I_q\nonumber\\
    =&\sum_{a=0}^{q-1}{A^{(a)}}^\dagger A^{(a)}\bigotimes I_q=I_\chi\bigotimes I_q.
\end{align}
Here the equation $\sum_{a=0}^{q-1}{A^{(a)}}^\dagger A^{(a)}=I_\chi$ comes from the left-canonical condition.

Similarly, for the case of two-site shift invariant initial MPS described in Sec. \ref{sec:twosite}, the Kraus operators are given by
\begin{equation}
    K_{a,a'}=\sum_{b=0}^{q-1} A^{(b)}B^{(a)}\bigotimes(\ket{b}\bra{a'})_{x=0}.
\end{equation}

\subsection{Mixed initial states}
Here, we generalize the quantum channel expression to the mixed initial states admitting Matrix Product Density Operator (MPDO) representation.
We further impose the locally purified condition, that is, the MPDO can be purified to a MPS with a local purifying auxiliary system \cite{Verstraete2004Matrix,Cuevas2013Purification}. MPDO with the locally purified condition is also termed as the Locally Purified Density Operator (LPDO). In this case, the three-leg local tensor $\mathcal{A}$ reads
\begin{equation}
\tikzset{every picture/.style={line width=0.75pt}} 
\begin{tikzpicture}[x=0.75pt,y=0.75pt,yscale=-1,xscale=1]

\draw [line width=1.5]    (120,143.04) -- (149,143.04) ;
\draw  [fill={rgb, 255:red, 208; green, 2; blue, 27 }  ,fill opacity=1 ] (128.47,135.68) .. controls (128.47,134.42) and (129.49,133.4) .. (130.75,133.4) -- (137.59,133.4) .. controls (138.85,133.4) and (139.87,134.42) .. (139.87,135.68) -- (139.87,150.39) .. controls (139.87,151.65) and (138.85,152.67) .. (137.59,152.67) -- (130.75,152.67) .. controls (129.49,152.67) and (128.47,151.65) .. (128.47,150.39) -- cycle ;
\draw  [line width=0.75]  (133.3,135.63) -- (137.3,135.63) -- (137.3,139.63) ;

\draw [fill={rgb, 255:red, 144; green, 19; blue, 254 }  ,fill opacity=1 ][line width=0.75]    (134.2,124) -- (134.25,132.9) ;

\draw (124.8,112.21) node [anchor=north west][inner sep=0.75pt]  [font=\scriptsize]  {$a,a'$};
\draw (101.4,135.61) node [anchor=north west][inner sep=0.75pt]  [font=\scriptsize]  {$j,j'$};
\draw (150.4,136.61) node [anchor=north west][inner sep=0.75pt]  [font=\scriptsize]  {$k,k'$};
\draw (172,129.7) node [anchor=north west][inner sep=0.75pt]    {$=\mathcal{A}_{jk,j'k'}^{( a,a')} =\sum _{\gamma =0}^{D-1} A_{jk}^{( a,\gamma )} A{_{j'k'}^{( a',\gamma )}}^{*}$};
\end{tikzpicture},
\end{equation}
where $\gamma$ is the auxiliary index corresponding to the local purification space of dimension $D$. In this context, the left-canonical condition is formulated as
\begin{equation}
   \sum_{\gamma=0}^{D-1}\sum_{a=0}^{q-1} {A^{(a,\gamma )}}^\dagger A^{(a,\gamma )}=I_\chi.
\end{equation}
For the $D=1$ case, the LPDO is reduced to the pure-state MPS studied in the main text. 

The LPDO corresponds to the quantum channel
\begin{equation}
    \mathcal{M}(\tilde{\rho}_R)=\sum_{\gamma'=0}^{D-1}\sum_{b,b'=0}^{q-1}A^{(b,\gamma')}(\sum_{\gamma=0}^{D-1}\sum_{a=0}^{q-1}A^{(a,\gamma)}\text{Tr}_{x=0}[\tilde{\rho}_R]{A^{(a,\gamma)}}^\dagger){A^{(b',\gamma')}}^\dagger\bigotimes (\ket{b}\bra{b'})_{x=0}.
\end{equation}
Following the similar analysis, we can extract the Kraus operators from the quantum channel expression:
\begin{equation}
    K_{(a\gamma,a'\gamma')}=\sum_{b=0}^{q-1} A^{(b,\gamma')}A^{(a,\gamma)}\bigotimes(\ket{b}\bra{a'})_{x=0} .
\end{equation}

\section{R$\acute{\text{E}}$nyi entropies dynamics}\label{sec:renyi}

Here, we preform tensor-network contractions to obtain analytical expressions for the time-evolved R$\acute{\text{e}}$nyi entropies $S^{(n)}_R(t)=\ln(\text{Tr}[\rho_R^n(t)])/(1-n)$, where $n\in\mathbb{Z}$ and $n>1$. We work in the space-time regime where size(L) and size(R) are larger than $2t$, such that the most efficient contractions of tensor networks from both sides are allowed. This regime corresponds to the early-stage growth of entanglement. Due to the presence of the strict light cone in brickwork architectures, we can extend size(L) and size(R) to be infinite, without altering the results.
As demonstrated in the main text, we impose two solvable conditions with opposite chirality, quoted as below:
\begin{equation}\label{eq:twofolded}
\tikzset{every picture/.style={line width=0.75pt}} 
\begin{tikzpicture}[x=0.75pt,y=0.75pt,yscale=-1,xscale=1]

\draw  [fill={rgb, 255:red, 144; green, 19; blue, 254 }  ,fill opacity=1 ][line width=1.5]  (208.33,82.29) .. controls (208.33,80.38) and (209.88,78.83) .. (211.79,78.83) -- (219.87,78.83) .. controls (221.78,78.83) and (223.33,80.38) .. (223.33,82.29) -- (223.33,90.37) .. controls (223.33,92.28) and (221.78,93.83) .. (219.87,93.83) -- (211.79,93.83) .. controls (209.88,93.83) and (208.33,92.28) .. (208.33,90.37) -- cycle ;
\draw  [fill={rgb, 255:red, 144; green, 19; blue, 254 }  ,fill opacity=1 ][line width=0.75]  (215.83,82.33) -- (219.83,82.33) -- (219.83,86.33) ;
\draw [fill={rgb, 255:red, 144; green, 19; blue, 254 }  ,fill opacity=1 ][line width=0.75]    (223.13,93.45) -- (233.33,103.83) ;
\draw [fill={rgb, 255:red, 144; green, 19; blue, 254 }  ,fill opacity=1 ][line width=0.75]    (198.33,68.83) -- (208.83,79.33) ;
\draw [fill={rgb, 255:red, 144; green, 19; blue, 254 }  ,fill opacity=1 ][line width=0.75]    (234,68.83) -- (223.11,79.78) ;
\draw [fill={rgb, 255:red, 144; green, 19; blue, 254 }  ,fill opacity=1 ][line width=0.75]    (209,92.5) -- (198.33,103.83) ;

\draw  [fill={rgb, 255:red, 255; green, 255; blue, 255 }  ,fill opacity=1 ] (195.53,68.83) .. controls (195.53,67.29) and (196.79,66.03) .. (198.33,66.03) .. controls (199.88,66.03) and (201.13,67.29) .. (201.13,68.83) .. controls (201.13,70.38) and (199.88,71.63) .. (198.33,71.63) .. controls (196.79,71.63) and (195.53,70.38) .. (195.53,68.83) -- cycle ;
\draw [line width=1.5]    (188.5,61.94) -- (188.57,120.12) ;
\draw  [fill={rgb, 255:red, 208; green, 2; blue, 27 }  ,fill opacity=1 ] (195.89,100.33) .. controls (197.15,100.33) and (198.17,101.35) .. (198.17,102.61) -- (198.17,109.45) .. controls (198.17,110.71) and (197.15,111.73) .. (195.89,111.73) -- (181.18,111.73) .. controls (179.92,111.73) and (178.9,110.71) .. (178.9,109.45) -- (178.9,102.61) .. controls (178.9,101.35) and (179.92,100.33) .. (181.18,100.33) -- cycle ;
\draw  [line width=0.75]  (195.94,105.17) -- (195.94,109.17) -- (191.94,109.17) ;

\draw [line width=1.5]    (267.5,61.94) -- (267.57,120.12) ;
\draw  [fill={rgb, 255:red, 208; green, 2; blue, 27 }  ,fill opacity=1 ] (274.89,71.33) .. controls (276.15,71.33) and (277.17,72.35) .. (277.17,73.61) -- (277.17,80.45) .. controls (277.17,81.71) and (276.15,82.73) .. (274.89,82.73) -- (260.18,82.73) .. controls (258.92,82.73) and (257.9,81.71) .. (257.9,80.45) -- (257.9,73.61) .. controls (257.9,72.35) and (258.92,71.33) .. (260.18,71.33) -- cycle ;
\draw  [line width=0.75]  (274.94,76.17) -- (274.94,80.17) -- (270.94,80.17) ;

\draw    (277.5,76.13) -- (299,76.12) ;
\draw [fill={rgb, 255:red, 255; green, 255; blue, 255 }  ,fill opacity=1 ]   (277.33,103.83) -- (299.83,103.82) ;
\draw  [fill={rgb, 255:red, 255; green, 255; blue, 255 }  ,fill opacity=1 ] (274.53,103.83) .. controls (274.53,102.29) and (275.79,101.03) .. (277.33,101.03) .. controls (278.88,101.03) and (280.13,102.29) .. (280.13,103.83) .. controls (280.13,105.38) and (278.88,106.63) .. (277.33,106.63) .. controls (275.79,106.63) and (274.53,105.38) .. (274.53,103.83) -- cycle ;
\draw  [fill={rgb, 255:red, 144; green, 19; blue, 254 }  ,fill opacity=1 ][line width=1.5]  (345.33,82.29) .. controls (345.33,80.38) and (346.88,78.83) .. (348.79,78.83) -- (356.87,78.83) .. controls (358.78,78.83) and (360.33,80.38) .. (360.33,82.29) -- (360.33,90.37) .. controls (360.33,92.28) and (358.78,93.83) .. (356.87,93.83) -- (348.79,93.83) .. controls (346.88,93.83) and (345.33,92.28) .. (345.33,90.37) -- cycle ;
\draw  [fill={rgb, 255:red, 144; green, 19; blue, 254 }  ,fill opacity=1 ][line width=0.75]  (352.83,82.33) -- (356.83,82.33) -- (356.83,86.33) ;
\draw [fill={rgb, 255:red, 144; green, 19; blue, 254 }  ,fill opacity=1 ][line width=0.75]    (360.13,93.45) -- (370.33,103.83) ;
\draw [fill={rgb, 255:red, 144; green, 19; blue, 254 }  ,fill opacity=1 ][line width=0.75]    (335.33,68.83) -- (345.83,79.33) ;
\draw [fill={rgb, 255:red, 144; green, 19; blue, 254 }  ,fill opacity=1 ][line width=0.75]    (371,68.83) -- (360.11,79.78) ;
\draw [fill={rgb, 255:red, 144; green, 19; blue, 254 }  ,fill opacity=1 ][line width=0.75]    (346,92.5) -- (335.33,103.83) ;

\draw  [fill={rgb, 255:red, 255; green, 255; blue, 255 }  ,fill opacity=1 ] (368.2,68.83) .. controls (368.2,67.29) and (369.45,66.03) .. (371,66.03) .. controls (372.55,66.03) and (373.8,67.29) .. (373.8,68.83) .. controls (373.8,70.38) and (372.55,71.63) .. (371,71.63) .. controls (369.45,71.63) and (368.2,70.38) .. (368.2,68.83) -- cycle ;
\draw [line width=1.5]    (380.5,59.94) -- (380.57,119.12) ;
\draw  [fill={rgb, 255:red, 208; green, 2; blue, 27 }  ,fill opacity=1 ] (372.18,110.73) .. controls (370.92,110.73) and (369.9,109.71) .. (369.9,108.45) -- (369.9,101.61) .. controls (369.9,100.35) and (370.92,99.33) .. (372.18,99.33) -- (386.89,99.33) .. controls (388.15,99.33) and (389.17,100.35) .. (389.17,101.61) -- (389.17,108.45) .. controls (389.17,109.71) and (388.15,110.73) .. (386.89,110.73) -- cycle ;
\draw  [line width=0.75]  (372.13,105.9) -- (372.13,101.9) -- (376.13,101.9) ;

\draw [line width=1.5]    (441.6,120.05) -- (441.6,61.87) ;
\draw  [fill={rgb, 255:red, 208; green, 2; blue, 27 }  ,fill opacity=1 ] (434.15,80.71) .. controls (432.89,80.72) and (431.86,79.7) .. (431.85,78.44) -- (431.81,71.6) .. controls (431.8,70.34) and (432.81,69.32) .. (434.07,69.31) -- (448.78,69.21) .. controls (450.04,69.2) and (451.06,70.21) .. (451.07,71.47) -- (451.12,78.31) .. controls (451.13,79.57) and (450.12,80.6) .. (448.86,80.61) -- cycle ;
\draw  [line width=0.75]  (434.07,75.88) -- (434.04,71.88) -- (438.04,71.85) ;

\draw    (431.5,75.1) -- (411,75.09) ;
\draw [fill={rgb, 255:red, 255; green, 255; blue, 255 }  ,fill opacity=1 ]   (433.63,104.3) -- (411.13,104.31) ;
\draw  [fill={rgb, 255:red, 255; green, 255; blue, 255 }  ,fill opacity=1 ] (436.43,104.57) .. controls (436.42,106.12) and (435.15,107.36) .. (433.6,107.34) .. controls (432.06,107.32) and (430.82,106.06) .. (430.83,104.51) .. controls (430.85,102.96) and (432.12,101.72) .. (433.66,101.74) .. controls (435.21,101.76) and (436.45,103.02) .. (436.43,104.57) -- cycle ;

\draw (234,82.74) node [anchor=north west][inner sep=0.75pt]    {$=$};
\draw (389.8,83.14) node [anchor=north west][inner sep=0.75pt]    {$=$};
\draw (312.8,104) node [anchor=north west][inner sep=0.75pt]    {$,$};
\draw (456.8,104) node [anchor=north west][inner sep=0.75pt]    {$.$};
\end{tikzpicture}
\end{equation}
Meanwhile, we choose the initial state as the homogeneous, one-site shift invariant MPS with local tensor $A$. To dispense with the influence of boundary conditions, we further assume that the MPS is in the left- and right-canonical form: $\sum_{a=0}^{q-1}{A^{(a)}}^\dagger A^{(a)}=\sum_{a=0}^{q-1}A^{(a)}{A^{(a)}}^\dagger=I_\chi$.

In Sec. \ref{sec:multi_folded}, we introduce basic notations of the multi-folded representation, which are demanded to present $\text{Tr}[\rho_R^n(t))]$ graphically. Subsequently, we apply the solvable conditions to contract the tensor network in Sec. \ref{sec:quantum_channel}, and establish the correspondence between the $n$th R$\acute{\text{e}}$nyi entropy at the time step $t$ and the repeated $2t$-time action of a quantum channel. 
Thus, this correspondence provides an operational approach to accurately calculating the initial growing rate of the entanglement. 
Sec. \ref{sec:purification} is devoted to demonstrating the contracted tensor network of $\text{Tr}[\rho_R^n(t))]$ as the $n$th R$\acute{\text{e}}$nyi entropy of a $4t$-length pure MPS state subjected to a special non-contiguous bipartition. Accordingly, the time-volved von Neumann entanglement entropy obtained through the analytic continuation $S_{\text{ent}}(t)=\lim_{n\to 1}\ln(\text{Tr}[\rho_R^n(t))])/(1-n)$ can be demonstrated as the $S_{\text{ent}}$ of this pure MPS state subjected to the same bipartition. We put examples of unitary gates allowing for exact calculations of R$\acute{\text{e}}$nyi entropy dynamics in Sec. \ref{sec:opposite}.

\subsection{Multi-folded representations}\label{sec:multi_folded}
In the main text, we have adopted the folded representation to represent the time-evolved state by a single patch of tensor network, i.e. Fig. 1(a) in the main text. 
Here, in order to represent the $n$th power of the state, we introduce the $n$-folded representation for unitary gates and MPS as follows:
\begin{equation}
\tikzset{every picture/.style={line width=0.75pt}} 
.
\end{equation}
Consequently, we can represent the $n-$replica of a time-evolved pure state as below, $t=2$ for example:
\begin{equation}
\tikzset{every picture/.style={line width=0.75pt}} 
%
\end{equation}

Then we trace out the degrees of freedom in L to obtain $\rho_R(t)^{\otimes n}$. In the $n-$folded representation, the tracing operation manifests as a pairing rule on each site: $\delta_{a_1,a'_1}\cdots \delta_{a_n,a'_n}$, where $a_m$ and $a'_m$ respectively correspond to the forward and backward branches on a single-site Hilbert space in the $m$th replica. We still use the hollow dot to represent such a pairing operator. The $n$-replica reduced
density matrix is represented as:
\begin{equation}
\tikzset{every picture/.style={line width=0.75pt}} 

\end{equation}

The next step is to insert a permutation operator $\mathbb{P}_n$ and trace out R: $\text{Tr}[\rho_R^n(t)]=\text{Tr}[\mathbb{P}_n\rho _{R}(t)^{\otimes n}]$ \cite{Bertini2023Exact}. Notably, this operation corresponds to conducting the following pairing rule on each site: $\delta_{a'_1,a_2}\delta_{a'_2,a_3}\cdots \delta_{a'_{n-1},a_n}\delta_{a'_n,a_1}$. We use the hollow diamond to represent this pairing operator \cite{Bertini2020Scrambling}:
\begin{equation}
\tikzset{every picture/.style={line width=0.75pt}} 

\end{equation}
Therefore, we have:
\begin{equation}\label{eq:folded_Renyi}
\tikzset{every picture/.style={line width=0.75pt}} 
%
\end{equation}

\subsection{Entanglement velocity}\label{sec:quantum_channel}
In this subsection, we show that the tensor network in Eq. \eqref{eq:folded_Renyi} can be efficiently contracted utilizing the two solvable conditions. The crucial point is to establish the contraction rules in the $n$-folded representation. 
First, notice that the time-unitarity condition leads to the following two rules:
\begin{equation}
\tikzset{every picture/.style={line width=0.75pt}} 
%
.
\end{equation}
The two contraction rules can be verified by noticing that both two kinds of pairing operators couple between $U$ and $U^*$. The unitarity together with left- and right-canonical property of the MPS allow contractions of the tensor network within the lightcone:
\begin{equation}
\tikzset{every picture/.style={line width=0.75pt}} 
%
\end{equation}

Second, and most importantly, notice that the solvable conditions can be represented in the $n$-folded picture exactly the same as those in the folded picture [Eq. \eqref{eq:twofolded}]:
\begin{equation}
\tikzset{every picture/.style={line width=0.75pt}} 
%
\end{equation}
Consequently, we obtain a highly simplified graphical representation:
\begin{equation}\label{eq:renyi1d}
\tikzset{every picture/.style={line width=0.75pt}} 
%
\end{equation}

The resulting one-dimensional tensor network can be demonstrated as the consecutive action of a two-site transfer matrix $\mathbb{T}$ with appropriate initial and finial states. Here $\mathbb{T}$ acts on $2n$-replica Hilbert space, defined as
\begin{equation}
\tikzset{every picture/.style={line width=0.75pt}} 
%
,
\end{equation}
and thus
\begin{equation}\label{eq:RenyiFormula}
    \text{Tr}\left[\rho _{R}^{n}(t)\right]=\langle \bullet |\mathbb{T}^{2t}|\soliddiamond\rangle,
\end{equation}
here $|\bullet\rangle$ and $|\soliddiamond\rangle$ correspond to the vectorizations of two different pairing operators in the folded auxiliary Hilbert space. Indeed, $\mathbb{T}$ realizes a completely positive quantum channel when mapped to be a superoperator in the $n$-replica Hilbert space, while the tracing-preserving property is not guaranteed. In the large time limit (but still smaller than the size of L and R), Eq. \eqref{eq:RenyiFormula} is dominated by the largest eigenvalue $\lambda_n$ (in the modulus) of the quantum channel $\mathbb{T}$, given by $\sim\lambda_n^{2t}$. 
Therefore, the $n$th entanglement velocity characterizing the asymptotic growth of the $n$th R$\acute{\text{e}}$nyi entropy is exactly determined by $\lambda_n$ as follows:
\begin{equation}
    v_E^{(n)}\equiv\lim_{t\to\infty}\frac{\ln(\text{Tr}[\rho_R^n(t)])/(1-n)}{t\ln(q)}=\frac{2\ln(\lambda_n)}{(1-n)\ln(q)}.
\end{equation}
Thus, this result adds a new example to exactly solvable entanglement dynamics. The analytic continuation towards $n=1$ demands more knowledge on how $\lambda_n$ depends on $n$.

\subsection{Entanglement of the temporal state}\label{sec:purification}
In this subsection, we will uncover a unique space-time duality between the entanglement of the time-evolved state under a contiguous bipartition, and that of a pure MPS state defined on the temporal Hilbert space under a non-contiguous bipartition.
More specifically, we will show that Eq. \eqref{eq:renyi1d} can be demonstrated as the $n$th R$\acute{\text{e}}$nyi entropy of a $(4t+1)$-length MPS state between all the even and odd sites.

We prove this statement as below. Consider the following pure state $\ket{\phi}$ defined in the Hilbert space $(\mathcal{H}_q^{\otimes 4t})\otimes \mathcal{H}_\chi$:
\begin{equation}
\tikzset{every picture/.style={line width=0.75pt}} 
\begin{tikzpicture}[x=0.75pt,y=0.75pt,yscale=-1,xscale=1]

\draw [fill={rgb, 255:red, 144; green, 19; blue, 254 }  ,fill opacity=1 ][line width=0.75]    (201.9,73.5) -- (201.87,85.36) ;
\draw  [fill={rgb, 255:red, 0; green, 0; blue, 0 }  ,fill opacity=1 ] (188.27,96.12) .. controls (186.72,96.12) and (185.47,94.86) .. (185.48,93.31) .. controls (185.48,91.77) and (186.74,90.52) .. (188.28,90.52) .. controls (189.83,90.52) and (191.08,91.78) .. (191.08,93.33) .. controls (191.07,94.87) and (189.81,96.12) .. (188.27,96.12) -- cycle ;
\draw [line width=1.5]    (187.93,93.54) -- (238.35,93.5) ;
\draw  [fill={rgb, 255:red, 208; green, 2; blue, 27 }  ,fill opacity=1 ] (197.04,85.4) .. controls (197.04,84.14) and (198.07,83.12) .. (199.33,83.13) -- (206.17,83.15) .. controls (207.43,83.15) and (208.44,84.17) .. (208.44,85.43) -- (208.4,100.14) .. controls (208.4,101.4) and (207.38,102.42) .. (206.12,102.41) -- (199.28,102.39) .. controls (198.02,102.39) and (197,101.37) .. (197,100.11) -- cycle ;
\draw  [fill={rgb, 255:red, 208; green, 2; blue, 27 }  ,fill opacity=1 ][line width=0.75]  (201.87,85.36) -- (205.87,85.37) -- (205.86,89.37) ;

\draw  [fill={rgb, 255:red, 208; green, 2; blue, 27 }  ,fill opacity=1 ] (216.31,85.46) .. controls (216.31,84.2) and (217.33,83.18) .. (218.59,83.19) -- (225.43,83.2) .. controls (226.69,83.21) and (227.71,84.23) .. (227.71,85.49) -- (227.67,100.2) .. controls (227.67,101.46) and (226.64,102.47) .. (225.38,102.47) -- (218.54,102.45) .. controls (217.28,102.45) and (216.27,101.43) .. (216.27,100.17) -- cycle ;
\draw  [fill={rgb, 255:red, 208; green, 2; blue, 27 }  ,fill opacity=1 ][line width=0.75]  (221.14,85.42) -- (225.14,85.43) -- (225.13,89.43) ;

\draw [fill={rgb, 255:red, 144; green, 19; blue, 254 }  ,fill opacity=1 ][line width=0.75]    (222.32,83.35) -- (222.57,73.5) ;
\draw [fill={rgb, 255:red, 144; green, 19; blue, 254 }  ,fill opacity=1 ][line width=0.75]    (279.57,73.5) -- (279.54,85.28) ;
\draw [line width=1.5]    (264.6,93.45) -- (314.02,93.41) ;
\draw  [fill={rgb, 255:red, 2; green, 208; blue, 21 }  ,fill opacity=1 ] (273.71,85.31) .. controls (273.71,84.05) and (274.73,83.04) .. (275.99,83.04) -- (282.83,83.06) .. controls (284.09,83.06) and (285.11,84.08) .. (285.11,85.34) -- (285.07,100.05) .. controls (285.07,101.31) and (284.04,102.33) .. (282.78,102.32) -- (275.94,102.31) .. controls (274.68,102.3) and (273.67,101.28) .. (273.67,100.02) -- cycle ;
\draw  [fill={rgb, 255:red, 2; green, 208; blue, 21 }  ,fill opacity=1 ][line width=0.75]  (278.54,85.28) -- (282.54,85.29) -- (282.53,89.29) ;

\draw  [fill={rgb, 255:red, 2; green, 208; blue, 21 }  ,fill opacity=1 ] (292.97,85.37) .. controls (292.98,84.11) and (294,83.1) .. (295.26,83.1) -- (302.1,83.12) .. controls (303.36,83.12) and (304.38,84.14) .. (304.37,85.4) -- (304.34,100.11) .. controls (304.33,101.37) and (303.31,102.39) .. (302.05,102.38) -- (295.21,102.37) .. controls (293.95,102.36) and (292.93,101.34) .. (292.94,100.08) -- cycle ;
\draw  [fill={rgb, 255:red, 2; green, 208; blue, 21 }  ,fill opacity=1 ][line width=0.75]  (297.81,85.33) -- (301.81,85.34) -- (301.8,89.34) ;

\draw [fill={rgb, 255:red, 144; green, 19; blue, 254 }  ,fill opacity=1 ][line width=0.75]    (298.99,83.26) -- (299.24,72.56) ;
\draw [line width=1.5]    (264.6,93.45) -- (314.02,93.41) ;
\draw  [fill={rgb, 255:red, 208; green, 2; blue, 27 }  ,fill opacity=1 ] (273.71,85.31) .. controls (273.71,84.05) and (274.73,83.04) .. (275.99,83.04) -- (282.83,83.06) .. controls (284.09,83.06) and (285.11,84.08) .. (285.11,85.34) -- (285.07,100.05) .. controls (285.07,101.31) and (284.04,102.33) .. (282.78,102.32) -- (275.94,102.31) .. controls (274.68,102.3) and (273.67,101.28) .. (273.67,100.02) -- cycle ;
\draw  [fill={rgb, 255:red, 208; green, 2; blue, 27 }  ,fill opacity=1 ][line width=0.75]  (278.54,85.28) -- (282.54,85.29) -- (282.53,89.29) ;

\draw  [fill={rgb, 255:red, 208; green, 2; blue, 27 }  ,fill opacity=1 ] (292.97,85.37) .. controls (292.98,84.11) and (294,83.1) .. (295.26,83.1) -- (302.1,83.12) .. controls (303.36,83.12) and (304.38,84.14) .. (304.37,85.4) -- (304.34,100.11) .. controls (304.33,101.37) and (303.31,102.39) .. (302.05,102.38) -- (295.21,102.37) .. controls (293.95,102.36) and (292.93,101.34) .. (292.94,100.08) -- cycle ;
\draw  [fill={rgb, 255:red, 208; green, 2; blue, 27 }  ,fill opacity=1 ][line width=0.75]  (297.81,85.33) -- (301.81,85.34) -- (301.8,89.34) ;

\draw [fill={rgb, 255:red, 144; green, 19; blue, 254 }  ,fill opacity=1 ][line width=0.75]    (298.99,83.26) -- (299.24,73.5) ;
\draw [fill={rgb, 255:red, 144; green, 19; blue, 254 }  ,fill opacity=1 ][line width=0.75]    (240.9,73.5) -- (240.87,85.36) ;
\draw [line width=1.5]    (228.93,93.54) -- (273.35,93.5) ;
\draw  [fill={rgb, 255:red, 208; green, 2; blue, 27 }  ,fill opacity=1 ] (236.04,85.4) .. controls (236.04,84.14) and (237.07,83.12) .. (238.33,83.13) -- (245.17,83.15) .. controls (246.43,83.15) and (247.44,84.17) .. (247.44,85.43) -- (247.4,100.14) .. controls (247.4,101.4) and (246.38,102.42) .. (245.12,102.41) -- (238.28,102.39) .. controls (237.02,102.39) and (236,101.37) .. (236,100.11) -- cycle ;
\draw  [fill={rgb, 255:red, 208; green, 2; blue, 27 }  ,fill opacity=1 ][line width=0.75]  (240.87,85.36) -- (244.87,85.37) -- (244.86,89.37) ;

\draw  [fill={rgb, 255:red, 208; green, 2; blue, 27 }  ,fill opacity=1 ] (255.31,85.46) .. controls (255.31,84.2) and (256.33,83.18) .. (257.59,83.19) -- (264.43,83.2) .. controls (265.69,83.21) and (266.71,84.23) .. (266.71,85.49) -- (266.67,100.2) .. controls (266.67,101.46) and (265.64,102.47) .. (264.38,102.47) -- (257.54,102.45) .. controls (256.28,102.45) and (255.27,101.43) .. (255.27,100.17) -- cycle ;
\draw  [fill={rgb, 255:red, 208; green, 2; blue, 27 }  ,fill opacity=1 ][line width=0.75]  (260.14,85.42) -- (264.14,85.43) -- (264.13,89.43) ;

\draw [fill={rgb, 255:red, 144; green, 19; blue, 254 }  ,fill opacity=1 ][line width=0.75]    (261.6,83.35) -- (261.57,73.5) ;
\draw [fill={rgb, 255:red, 144; green, 19; blue, 254 }  ,fill opacity=1 ][line width=0.75]    (316.23,73.5) -- (316.21,85.36) ;
\draw [line width=1.5]    (304.26,93.54) -- (351.69,93.5) -- (351.7,74.27) ;
\draw  [fill={rgb, 255:red, 208; green, 2; blue, 27 }  ,fill opacity=1 ] (310.71,85.65) .. controls (310.71,84.39) and (311.73,83.37) .. (312.99,83.37) -- (319.83,83.39) .. controls (321.09,83.39) and (322.11,84.42) .. (322.11,85.68) -- (322.07,100.38) .. controls (322.07,101.64) and (321.04,102.66) .. (319.78,102.66) -- (312.94,102.64) .. controls (311.68,102.64) and (310.67,101.61) .. (310.67,100.35) -- cycle ;
\draw  [fill={rgb, 255:red, 208; green, 2; blue, 27 }  ,fill opacity=1 ][line width=0.75]  (315.54,85.61) -- (319.54,85.62) -- (319.53,89.62) ;

\draw  [fill={rgb, 255:red, 208; green, 2; blue, 27 }  ,fill opacity=1 ] (330.64,85.46) .. controls (330.64,84.2) and (331.67,83.18) .. (332.93,83.19) -- (339.77,83.2) .. controls (341.03,83.21) and (342.04,84.23) .. (342.04,85.49) -- (342,100.2) .. controls (342,101.46) and (340.98,102.47) .. (339.72,102.47) -- (332.88,102.45) .. controls (331.62,102.45) and (330.6,101.43) .. (330.6,100.17) -- cycle ;
\draw  [fill={rgb, 255:red, 208; green, 2; blue, 27 }  ,fill opacity=1 ][line width=0.75]  (335.47,85.42) -- (339.47,85.43) -- (339.46,89.43) ;

\draw [fill={rgb, 255:red, 144; green, 19; blue, 254 }  ,fill opacity=1 ][line width=0.75]    (336.47,83.42) -- (336.72,73.5) ;

\draw (124,85.56) node [anchor=north west][inner sep=0.75pt]    {$|\phi \rangle \langle \phi |=$};

\end{tikzpicture}
\end{equation}
We bipartite the Hilbert space to two non-contiguous regions $E$ and $O$, while even (odd) sites belong to $E$ ($O$), respectively. The reduced density matrix of $\ket{\phi}$ on $O$ is:
\begin{equation}
\tikzset{every picture/.style={line width=0.75pt}} 
\begin{tikzpicture}[x=0.75pt,y=0.75pt,yscale=-1,xscale=1]

\draw [fill={rgb, 255:red, 144; green, 19; blue, 254 }  ,fill opacity=1 ][line width=0.75]    (201.9,73.5) -- (201.87,85.36) ;
\draw  [fill={rgb, 255:red, 0; green, 0; blue, 0 }  ,fill opacity=1 ] (188.27,96.12) .. controls (186.72,96.12) and (185.47,94.86) .. (185.48,93.31) .. controls (185.48,91.77) and (186.74,90.52) .. (188.28,90.52) .. controls (189.83,90.52) and (191.08,91.78) .. (191.08,93.33) .. controls (191.07,94.87) and (189.81,96.12) .. (188.27,96.12) -- cycle ;
\draw [line width=1.5]    (187.93,93.54) -- (238.35,93.5) ;
\draw  [fill={rgb, 255:red, 208; green, 2; blue, 27 }  ,fill opacity=1 ] (197.04,85.4) .. controls (197.04,84.14) and (198.07,83.12) .. (199.33,83.13) -- (206.17,83.15) .. controls (207.43,83.15) and (208.44,84.17) .. (208.44,85.43) -- (208.4,100.14) .. controls (208.4,101.4) and (207.38,102.42) .. (206.12,102.41) -- (199.28,102.39) .. controls (198.02,102.39) and (197,101.37) .. (197,100.11) -- cycle ;
\draw  [fill={rgb, 255:red, 208; green, 2; blue, 27 }  ,fill opacity=1 ][line width=0.75]  (201.87,85.36) -- (205.87,85.37) -- (205.86,89.37) ;

\draw  [fill={rgb, 255:red, 208; green, 2; blue, 27 }  ,fill opacity=1 ] (216.31,85.46) .. controls (216.31,84.2) and (217.33,83.18) .. (218.59,83.19) -- (225.43,83.2) .. controls (226.69,83.21) and (227.71,84.23) .. (227.71,85.49) -- (227.67,100.2) .. controls (227.67,101.46) and (226.64,102.47) .. (225.38,102.47) -- (218.54,102.45) .. controls (217.28,102.45) and (216.27,101.43) .. (216.27,100.17) -- cycle ;
\draw  [fill={rgb, 255:red, 208; green, 2; blue, 27 }  ,fill opacity=1 ][line width=0.75]  (221.14,85.42) -- (225.14,85.43) -- (225.13,89.43) ;

\draw [fill={rgb, 255:red, 144; green, 19; blue, 254 }  ,fill opacity=1 ][line width=0.75]    (222.32,83.35) -- (222.57,73.5) ;
\draw [fill={rgb, 255:red, 144; green, 19; blue, 254 }  ,fill opacity=1 ][line width=0.75]    (279.57,73.5) -- (279.54,85.28) ;
\draw [line width=1.5]    (264.6,93.45) -- (314.02,93.41) ;
\draw  [fill={rgb, 255:red, 2; green, 208; blue, 21 }  ,fill opacity=1 ] (273.71,85.31) .. controls (273.71,84.05) and (274.73,83.04) .. (275.99,83.04) -- (282.83,83.06) .. controls (284.09,83.06) and (285.11,84.08) .. (285.11,85.34) -- (285.07,100.05) .. controls (285.07,101.31) and (284.04,102.33) .. (282.78,102.32) -- (275.94,102.31) .. controls (274.68,102.3) and (273.67,101.28) .. (273.67,100.02) -- cycle ;
\draw  [fill={rgb, 255:red, 2; green, 208; blue, 21 }  ,fill opacity=1 ][line width=0.75]  (278.54,85.28) -- (282.54,85.29) -- (282.53,89.29) ;

\draw  [fill={rgb, 255:red, 2; green, 208; blue, 21 }  ,fill opacity=1 ] (292.97,85.37) .. controls (292.98,84.11) and (294,83.1) .. (295.26,83.1) -- (302.1,83.12) .. controls (303.36,83.12) and (304.38,84.14) .. (304.37,85.4) -- (304.34,100.11) .. controls (304.33,101.37) and (303.31,102.39) .. (302.05,102.38) -- (295.21,102.37) .. controls (293.95,102.36) and (292.93,101.34) .. (292.94,100.08) -- cycle ;
\draw  [fill={rgb, 255:red, 2; green, 208; blue, 21 }  ,fill opacity=1 ][line width=0.75]  (297.81,85.33) -- (301.81,85.34) -- (301.8,89.34) ;

\draw [fill={rgb, 255:red, 144; green, 19; blue, 254 }  ,fill opacity=1 ][line width=0.75]    (298.99,83.26) -- (299.24,72.56) ;
\draw [line width=1.5]    (264.6,93.45) -- (314.02,93.41) ;
\draw  [fill={rgb, 255:red, 208; green, 2; blue, 27 }  ,fill opacity=1 ] (273.71,85.31) .. controls (273.71,84.05) and (274.73,83.04) .. (275.99,83.04) -- (282.83,83.06) .. controls (284.09,83.06) and (285.11,84.08) .. (285.11,85.34) -- (285.07,100.05) .. controls (285.07,101.31) and (284.04,102.33) .. (282.78,102.32) -- (275.94,102.31) .. controls (274.68,102.3) and (273.67,101.28) .. (273.67,100.02) -- cycle ;
\draw  [fill={rgb, 255:red, 208; green, 2; blue, 27 }  ,fill opacity=1 ][line width=0.75]  (278.54,85.28) -- (282.54,85.29) -- (282.53,89.29) ;

\draw  [fill={rgb, 255:red, 208; green, 2; blue, 27 }  ,fill opacity=1 ] (292.97,85.37) .. controls (292.98,84.11) and (294,83.1) .. (295.26,83.1) -- (302.1,83.12) .. controls (303.36,83.12) and (304.38,84.14) .. (304.37,85.4) -- (304.34,100.11) .. controls (304.33,101.37) and (303.31,102.39) .. (302.05,102.38) -- (295.21,102.37) .. controls (293.95,102.36) and (292.93,101.34) .. (292.94,100.08) -- cycle ;
\draw  [fill={rgb, 255:red, 208; green, 2; blue, 27 }  ,fill opacity=1 ][line width=0.75]  (297.81,85.33) -- (301.81,85.34) -- (301.8,89.34) ;

\draw [fill={rgb, 255:red, 144; green, 19; blue, 254 }  ,fill opacity=1 ][line width=0.75]    (298.99,83.26) -- (299.24,73.5) ;
\draw [fill={rgb, 255:red, 144; green, 19; blue, 254 }  ,fill opacity=1 ][line width=0.75]    (240.9,73.5) -- (240.87,85.36) ;
\draw [line width=1.5]    (228.93,93.54) -- (273.35,93.5) ;
\draw  [fill={rgb, 255:red, 208; green, 2; blue, 27 }  ,fill opacity=1 ] (236.04,85.4) .. controls (236.04,84.14) and (237.07,83.12) .. (238.33,83.13) -- (245.17,83.15) .. controls (246.43,83.15) and (247.44,84.17) .. (247.44,85.43) -- (247.4,100.14) .. controls (247.4,101.4) and (246.38,102.42) .. (245.12,102.41) -- (238.28,102.39) .. controls (237.02,102.39) and (236,101.37) .. (236,100.11) -- cycle ;
\draw  [fill={rgb, 255:red, 208; green, 2; blue, 27 }  ,fill opacity=1 ][line width=0.75]  (240.87,85.36) -- (244.87,85.37) -- (244.86,89.37) ;

\draw  [fill={rgb, 255:red, 208; green, 2; blue, 27 }  ,fill opacity=1 ] (255.31,85.46) .. controls (255.31,84.2) and (256.33,83.18) .. (257.59,83.19) -- (264.43,83.2) .. controls (265.69,83.21) and (266.71,84.23) .. (266.71,85.49) -- (266.67,100.2) .. controls (266.67,101.46) and (265.64,102.47) .. (264.38,102.47) -- (257.54,102.45) .. controls (256.28,102.45) and (255.27,101.43) .. (255.27,100.17) -- cycle ;
\draw  [fill={rgb, 255:red, 208; green, 2; blue, 27 }  ,fill opacity=1 ][line width=0.75]  (260.14,85.42) -- (264.14,85.43) -- (264.13,89.43) ;

\draw [fill={rgb, 255:red, 144; green, 19; blue, 254 }  ,fill opacity=1 ][line width=0.75]    (261.6,83.35) -- (261.57,73.5) ;
\draw [fill={rgb, 255:red, 144; green, 19; blue, 254 }  ,fill opacity=1 ][line width=0.75]    (316.23,73.5) -- (316.21,85.36) ;
\draw [line width=1.5]    (304.26,93.54) -- (351.69,93.5) -- (351.7,74.27) ;
\draw  [fill={rgb, 255:red, 208; green, 2; blue, 27 }  ,fill opacity=1 ] (310.71,85.65) .. controls (310.71,84.39) and (311.73,83.37) .. (312.99,83.37) -- (319.83,83.39) .. controls (321.09,83.39) and (322.11,84.42) .. (322.11,85.68) -- (322.07,100.38) .. controls (322.07,101.64) and (321.04,102.66) .. (319.78,102.66) -- (312.94,102.64) .. controls (311.68,102.64) and (310.67,101.61) .. (310.67,100.35) -- cycle ;
\draw  [fill={rgb, 255:red, 208; green, 2; blue, 27 }  ,fill opacity=1 ][line width=0.75]  (315.54,85.61) -- (319.54,85.62) -- (319.53,89.62) ;

\draw  [fill={rgb, 255:red, 208; green, 2; blue, 27 }  ,fill opacity=1 ] (330.64,85.46) .. controls (330.64,84.2) and (331.67,83.18) .. (332.93,83.19) -- (339.77,83.2) .. controls (341.03,83.21) and (342.04,84.23) .. (342.04,85.49) -- (342,100.2) .. controls (342,101.46) and (340.98,102.47) .. (339.72,102.47) -- (332.88,102.45) .. controls (331.62,102.45) and (330.6,101.43) .. (330.6,100.17) -- cycle ;
\draw  [fill={rgb, 255:red, 208; green, 2; blue, 27 }  ,fill opacity=1 ][line width=0.75]  (335.47,85.42) -- (339.47,85.43) -- (339.46,89.43) ;

\draw [fill={rgb, 255:red, 144; green, 19; blue, 254 }  ,fill opacity=1 ][line width=0.75]    (336.47,83.42) -- (336.72,73.5) ;
\draw  [fill={rgb, 255:red, 255; green, 255; blue, 255 }  ,fill opacity=1 ] (222.57,76.3) .. controls (221.02,76.3) and (219.77,75.04) .. (219.77,73.49) .. controls (219.78,71.95) and (221.04,70.7) .. (222.58,70.7) .. controls (224.13,70.7) and (225.38,71.96) .. (225.37,73.51) .. controls (225.37,75.05) and (224.11,76.3) .. (222.57,76.3) -- cycle ;
\draw  [fill={rgb, 255:red, 255; green, 255; blue, 255 }  ,fill opacity=1 ] (336.72,76.3) .. controls (335.17,76.3) and (333.92,75.04) .. (333.92,73.49) .. controls (333.93,71.95) and (335.19,70.7) .. (336.73,70.7) .. controls (338.28,70.7) and (339.53,71.96) .. (339.52,73.51) .. controls (339.52,75.05) and (338.26,76.3) .. (336.72,76.3) -- cycle ;
\draw  [fill={rgb, 255:red, 255; green, 255; blue, 255 }  ,fill opacity=1 ] (299.23,75.36) .. controls (297.69,75.36) and (296.44,74.1) .. (296.44,72.56) .. controls (296.45,71.01) and (297.7,69.76) .. (299.25,69.76) .. controls (300.79,69.77) and (302.05,71.02) .. (302.04,72.57) .. controls (302.04,74.12) and (300.78,75.37) .. (299.23,75.36) -- cycle ;
\draw  [fill={rgb, 255:red, 255; green, 255; blue, 255 }  ,fill opacity=1 ] (261.57,76.3) .. controls (260.02,76.3) and (258.77,75.04) .. (258.77,73.49) .. controls (258.78,71.95) and (260.04,70.7) .. (261.58,70.7) .. controls (263.13,70.7) and (264.38,71.96) .. (264.37,73.51) .. controls (264.37,75.05) and (263.11,76.3) .. (261.57,76.3) -- cycle ;

\draw (145,88) node [anchor=north west][inner sep=0.75pt]    {$\rho _{O} =$};

\end{tikzpicture}
\end{equation}
Following the notations of $n$-folded representation introduced in Sec. \ref{sec:multi_folded}, it is direct to show that $\text{Tr}[\rho_O^n]$ is given by Eq. \eqref{eq:renyi1d}. Furthermore, the von Neumann entanglement entropy of $\ket{\Psi(t)}$ is also mapped to that of $\ket{\phi}$ through the analytic continuation:
\begin{equation}
    S_\text{ent}(t)=\text{Tr}[\rho_R(t)\ln(\rho_R(t))]=\text{Tr}[\rho_O\ln(\rho_O)].
\end{equation}

In addition, the space-time duality sets an upper bound for the entanglement velocity. The maximal entanglement of $\ket{\phi}$ is approached when the reduced density matrix is maximally mixed: $\rho_O=(\frac{1}{q}I_q)^{\otimes 2t}\otimes (\frac{1}{\chi}I_\chi)$, which leads to the velocity 
\begin{equation}
    \max \{v_E^{(n)}\}=\lim_{t\to\infty}\frac{2t\ln(q^{1-n})+\ln(\chi^{1-n})}{(1-n)t\ln(q)}=2,
\end{equation}
and thus saturates the light-cone velocity in the brickwork architecture.

\section{$q=2$ solutions for the solvable condition}
In this section we focus on solving the two-qubit gates fulfilling the solvable condition. The solutions for both $\tilde{q}=1$ and $\tilde{q}=2$ are covered by the dual-unitary gates \cite{Bertini2019Exact}. For $\tilde{q}=1$, we show that the solutions feature in chiral solitons \cite{Bertini2020OperatorII}, given by Eq. (11) in the main text (see also Sec. \ref{sec:single}). On the other hand, the $\tilde{q}=2$ solutions are {\footnotesize${\mathrm{SWAP}}$} gates dressed by single-site unitaries (Sec. \ref{sec:double}).

\subsection{Preliminary: two-qubit gates}
To begin with, notice that every two-qubit unitary gate can be parameterized as \cite{Khaneja2001Time,Kraus2001Optimal,Zhang2003Geometric}:
\begin{equation}\label{eq:universal}
    U=e^{i\phi}(u_+\otimes u_-)V[J_1,J_2,J_3](v_-\otimes v_+),
\end{equation}
where $u_\pm,v_\pm\in \text{SU}(2)$, and
\begin{equation}
    V[J_1,J_2,J_3]=\exp{[-i(J_1\sigma^1\otimes\sigma^1+J_2\sigma^2\otimes\sigma^2+J_3\sigma^3\otimes\sigma^3)]}.
\end{equation}
Since increasing any $J_\alpha$ by $\frac{\pi}{2}$ is equivalent to applying a tensor product of two single-site unitaries which can be absorbed into $u_\pm,v_\pm$, we can restrict the range of $J_\alpha$ within $[0,\frac{\pi}{2})$.
The matrix $V[J_1,J_2,J_3]$ can also be written on the Pauli basis \cite{Bertini2020OperatorII}:
\begin{equation}
V[J_1,J_2,J_3]=\sum_{\alpha=0}^{3}V_\alpha(J_1,J_2,J_3)\sigma^\alpha\otimes\sigma^\alpha,
\end{equation}
where the identity has been included in the Pauli matrices as $\sigma^0=I_2$.
Here $V_\alpha(J_1,J_2,J_3)$ are some complex-valued coefficients:
\begin{eqnarray}
    &V_0(J_1,J_2,J_3)=\cos(J_1)\cos(J_2)\cos(J_3)-i\sin(J_1)\sin(J_2)\sin(J_3),\nonumber\\
    &V_1(J_1,J_2,J_3)=\cos(J_1)\sin(J_2)\sin(J_3)-i\sin(J_1)\cos(J_2)\cos(J_3),\nonumber\\
    &V_2(J_1,J_2,J_3)=\sin(J_1)\cos(J_2)\sin(J_3)-i\cos(J_1)\sin(J_2)\cos(J_3),\nonumber\\
    &V_3(J_1,J_2,J_3)=\sin(J_1)\sin(J_2)\cos(J_3)-i\cos(J_1)\cos(J_2)\sin(J_3).
\end{eqnarray}
We adopt this parameterization and identify the constraints on all the parameters ($u_\pm,v_\pm$ and $J_{1,2,3}$) under the solvable condition.

\subsection{$q=2,\tilde{q}=1$}\label{sec:single}
Here we choose the one-site state to be $\ket{0}$, that is, $\mathcal{H}_A=\ket{0}\bra{0}$. The solvable condition reads $U^R(\ket{0}\bra{0}\otimes I_2)(U^R)^\dagger=I_2\otimes\ket{0}\bra{0}$, where $U^R$ represents the reshuffled operator of $U$ as defined in the main text.

In terms of Eq. \eqref{eq:universal}, the solvable condition can be rewritten as
\begin{equation}\label{eq:re_solvable}
    V^R[J_1,J_2,J_3][(v_-\ket{0}\bra{0}v_-^\dagger)\otimes I_2](V^R[J_1,J_2,J_3])^\dagger=I_2\otimes(u_-^\dagger\ket{0}\bra{0}u_-).
\end{equation}
We consider a \textit{necessary condition} for this equation by partially tracing out the first site degrees of freedom in this equation. To perform the trace operation, a trick is to rewrite the matrix $V[J_1,J_2,J_3]$ as
\begin{equation}
    V[J_1,J_2,J_3]=V[J'_1,J'_2,J'_3]V[\frac{\pi}{4},\frac{\pi}{4},\frac{\pi}{4}],
\end{equation}
where $V[\frac{\pi}{4},\frac{\pi}{4},\frac{\pi}{4}]$ is the {\footnotesize${\mathrm{SWAP}}$} gate exchaning the states of two qubits, $J'_\alpha=J_\alpha-\frac{\pi}{4}$, $-\frac{\pi}{4}\le J'_\alpha< \frac{\pi}{4}$. From now on, we will abbreviate the coefficients $V_\alpha(J'_1,J'_2,J'_3)$ to $V'_\alpha$. 
Next, we apply the partial trace on Eq. \eqref{eq:re_solvable} and write it in the time direction:
\begin{equation}\label{eq:rre_solvable}
    \text{Tr}_l\{V[J'_1,J'_2,J'_3][I_2\otimes(v_-\ket{0}\bra{0}v_-^\dagger)]V^\dagger[J'_1,J'_2,J'_3]\}=2 u_-^\dagger\ket{0}\bra{0}u_-.
\end{equation}
We can calculate this equation by expanding the expression in the curly brackets on the Pauli basis of the left site, followed by the trace operation. It leaves 
\begin{equation}
    \sum_{\alpha=0}^3|V'_\alpha|^2(\sigma^\alpha v_-\ket{0}\bra{0}v_-^\dagger \sigma^\alpha)=u_-^\dagger\ket{0}\bra{0}u_-.
\end{equation}
The left hand side is the summation of four rank-$1$ projectors with positive semi-definite coefficients, while on the right hand side there is one rank-$1$ projector. The equality holds only when the projector $u_-^\dagger\ket{0}\bra{0}u_-$ is exactly the same as those projectors $\sigma^\alpha v_-\ket{0}\bra{0}v_-^\dagger \sigma^\alpha$ with non-zero coefficients $|V'_\alpha|^2$.

We now show that it is impossible that three or all of $V'_\alpha$ are non-zero. Consider the case that $V'_\alpha$ and $V'_\beta$ are non-zero, and thus $\sigma^\alpha v_-\ket{0}$ and $\sigma^\beta v_-\ket{0}$ are the same as $u_-^\dagger\ket{0}$ up to a phase factor, $\alpha\neq\beta$. We can directly deduce that $v_-\ket{0}$ is an eigenstate of $\sigma^\alpha\sigma^\beta$. If there is another non-zero coefficient $V'_\gamma$, $v_-\ket{0}$ should be an eigenstate of $\sigma^\beta\sigma^\gamma$, which is impossible because $[\sigma^\alpha\sigma^\beta,\sigma^\beta\sigma^\gamma]\neq 0$. Therefore, the three coefficients $V'_{\alpha,\beta,\gamma}$ can not simultaneously be non-zero. Meanwhile, due to range of parameters $-\frac{\pi}{4}\le J'_\alpha< \frac{\pi}{4}$, $V'_0$ can never be zero. Hence, we only need to consider the following two classes:

\noindent (1) $V'_0$ and one of $V'_{\alpha\neq 0}$ are non-zero;

\noindent (2) Only $V'_0$ is non-zero.

We start from the first class and take $\alpha=3$ for instance. It follows that $v_-\ket{0}$ and $u_-^\dagger\ket{0}$ are eigenstates of $\sigma^3$ with the same eigenvalue. We can therefore parameterize the single-site special unitaries as: (I) $v_-=e^{-i\eta\sigma^3},u_-=e^{-i\epsilon\sigma^3}$, or (II) $v_-=i\sigma^1 e^{-i\eta\sigma^3},u_-=i e^{-i\epsilon\sigma^3}\sigma^1$. As for the coefficients, we have 
\begin{equation}
    |V'_0|^2+|V'_3|^2=1,V'_1=V'_2=0,
\end{equation}
of which the only solution is $J'_1=J'_2=0$. It can be checked that this solution combined with (I) fulfill the original solvable condition Eq. \eqref{eq:re_solvable}. This results in the set of unitary gates parameterized as Eq. (11) in the main text:
\begin{eqnarray}\label{eq:TheSolution}
    U&=&e^{i\phi}(u_+\otimes e^{-i\epsilon\sigma^3})V[0,0,J'_3]V[\frac{\pi}{4},\frac{\pi}{4},\frac{\pi}{4}](e^{-i\eta\sigma^3}\otimes v_+)\nonumber\\
    &=&e^{i\phi}(u_+\otimes e^{-i\epsilon\sigma^3})V[\frac{\pi}{4},\frac{\pi}{4},J_3](e^{-i\eta\sigma^3}\otimes v_+).
\end{eqnarray}
On the other hand, the single-site unitaries (II) can be included into Eq. \eqref{eq:TheSolution} through the following transformation:
\begin{eqnarray}
    U&=&e^{i\phi}[u_+\otimes (ie^{-i\epsilon\sigma^3}\sigma^1)]V[\frac{\pi}{4},\frac{\pi}{4},J_3][(i\sigma^1e^{-i\eta\sigma^3})\otimes v_+]\nonumber\\
    &=&e^{i\phi}[(iu_+\sigma^1)\otimes e^{-i\epsilon\sigma^3}]V[\frac{\pi}{4},\frac{\pi}{4},J_3][e^{-i\eta\sigma^3}\otimes(i\sigma^1v_+)].
\end{eqnarray}
Other cases in the first class ($\alpha=1$ or $2$) can be analyzed following a similar approach. It turns out that all the solutions are covered by Eq. \eqref{eq:TheSolution}.

Next, we consider the second class, where $V'_1=V'_2=V'_3=0$. This condition leads to $J'_\alpha=0$, and $v_-u_-=e^{i\alpha}$. The solutions read
\begin{equation}
    U=e^{i\phi}(u_+\otimes u_-)V[\frac{\pi}{4},\frac{\pi}{4},\frac{\pi}{4}](v_-\otimes v_+)=e^{i(\phi+\alpha)}(u_+\otimes I_2)V[\frac{\pi}{4},\frac{\pi}{4},\frac{\pi}{4}](I_2\otimes v_+),
\end{equation}
which, again, are covered by Eq. \eqref{eq:TheSolution}. 

To conclude, we have exhausted the two-qubit gates fulfilling the solvable condition for $\tilde{q}=1$. We obtain the solutions Eq. \eqref{eq:TheSolution} by rigorously classifying and discussing the consequences of a \textit{necessary condition} Eq. \eqref{eq:rre_solvable}, while the solutions can be verified to fulfill the \textit{sufficient condition}. We point out that all the gates given by Eq. \eqref{eq:rre_solvable} are dual-unitary gates \cite{Bertini2019Exact}.
Remarkably, the solutions coincide with those two-qubit circuits characterized by chiral solitons \cite{Bertini2020OperatorII}, defined by the soliton condition: $U^\dagger(\sigma^3\otimes I)U=I\otimes\sigma^3$.

\subsection{$q=2,\tilde{q}=2$}\label{sec:double}
In this subsection we consider $\tilde{q}=2$ where $\mathcal{H}_A=\mathcal{H}_{q=2}$. 
Notice that solutions for $\tilde{q}=2$ must form a subset of solutions for $\tilde{q}=1$.
Therefore, we can apply the $\tilde{q}=2$ solvable condition on $\tilde{q}=1$ solutions to further restrict the parameters. It follows that 
\begin{equation}
V^R[\frac{\pi}{4},\frac{\pi}{4},J_3](e^{-i\eta\sigma^3}\rho e^{i\eta\sigma^3}\otimes I_2)(V^R[\frac{\pi}{4},\frac{\pi}{4},J_3])^\dagger=I_2\otimes e^{i\epsilon\sigma^3}\rho e^{-i\epsilon\sigma^3},
\end{equation}
which holds for arbitrary one-site operator $\rho$.

We consider a sufficient condition by partially tracing out the first site degrees of freedom, and expand the matrix $V$ on the Pauli basis, which gives:
\begin{equation}\label{eq:channel_condition}
    \cos^2(J'_3)(e^{-i(\eta+\epsilon)\sigma^3}\rho e^{i(\eta+\epsilon)\sigma^3})+\sin^2(J'_3)(\sigma^3 e^{-i(\eta+\epsilon)\sigma^3}\rho e^{i(\eta+\epsilon)\sigma^3}\sigma^3)=\rho.
\end{equation}
The left hand side could be viewed as a quantum channel with two Kraus operators: $K_0=\cos(J'_3)e^{-i(\eta+\epsilon)\sigma^3},K_1=\sin(J'_3)\sigma^3e^{-i(\eta+\epsilon)\sigma^3}$. In this context, Eq. \eqref{eq:channel_condition} states that the quantum channel should be an identity operation, which requires that $J'_3=0, \eta+\epsilon=0$. The corresponding gates are 
\begin{eqnarray}\label{eq:The2Solution}
    U=e^{i\phi}(u\otimes I_2)S,
\end{eqnarray}
where $S=V[\frac{\pi}{4},\frac{\pi}{4},\frac{\pi}{4}]$ is the {\footnotesize${\mathrm{SWAP}}$} gate. It can be checked that all the gates in the form of Eq. \eqref{eq:The2Solution} fulfill the $\tilde{q}=2$ solvable condition.

\section{Solvable conditions with opposite chirality}\label{sec:opposite}
In this section, we discuss the relationship between two solvable conditions with opposite chirality [Eq. \eqref{eq:twofolded}]. We also present several examples that simultaneously fulfill both solvable conditions, allowing for exact calculations of  R$\acute{\text{e}}$nyi entropy dynamics, as demonstrated in Sec. \ref{sec:renyi}.

First of all, we establish a one-to-one correspondence between the solutions of two solvable conditions. To be concrete, given the three-leg tensor $A_{jk}^{(a)}$ as input, the solvable condition Eq. \eqref{eq:solvable} can be expressed as
\begin{equation}
U^R(\ket{A_{jk}}\bra{A_{j'k'}}\otimes I_q)(U^R)^\dagger=I_q\otimes\ket{A_{jk}}\bra{A_{j'k'}}.
\end{equation}
We show that $U_S=SUS$ ($S$ is the {\footnotesize${\mathrm{SWAP}}$} gate) must satisfy the opposite-chirality solvable condition. We denote the reshuffled operator of $U_S$ as $U_S^R$, and express the opposite-chirality solvable condition in the space direction from right to left, which involves the action of $(U_S^R)^T$ as:
\begin{equation}
    (U_S^R)^T(\ket{A_{jk}}\bra{A_{j'k'}}\otimes I_q)(U_S^R)^*=U_R(\ket{A_{jk}}\bra{A_{j'k'}}\otimes I_q)(U_R)^\dagger=I_q\otimes\ket{A_{jk}}\bra{A_{j'k'}}.
\end{equation}
Here, we have utilized the transformation $(U_S^R)^T=U^R$, which can be proven as follows \cite{Aravinda2021From}:
\begin{equation}
    [(U_S^R)^T]_{ab}^{cd}=(U_S^R)_{cd}^{ab}=(SUS)_{ca}^{db}=U_{bd}^{ac}=(U^R)_{ab}^{cd}.
\end{equation}
We illustrate the above analysis in the following diagrammatic representation:
\begin{equation}
\tikzset{every picture/.style={line width=0.75pt}} 
\begin{tikzpicture}[x=0.75pt,y=0.75pt,yscale=-1,xscale=1]

\draw    (297,46.83) .. controls (263.2,50.96) and (247.87,54.29) .. (271.2,61.29) ;
\draw  [fill={rgb, 255:red, 144; green, 19; blue, 254 }  ,fill opacity=1 ][line width=1.5]  (83.33,64.29) .. controls (83.33,62.38) and (84.88,60.83) .. (86.79,60.83) -- (94.87,60.83) .. controls (96.78,60.83) and (98.33,62.38) .. (98.33,64.29) -- (98.33,72.37) .. controls (98.33,74.28) and (96.78,75.83) .. (94.87,75.83) -- (86.79,75.83) .. controls (84.88,75.83) and (83.33,74.28) .. (83.33,72.37) -- cycle ;
\draw  [fill={rgb, 255:red, 144; green, 19; blue, 254 }  ,fill opacity=1 ][line width=0.75]  (90.83,64.33) -- (94.83,64.33) -- (94.83,68.33) ;
\draw [fill={rgb, 255:red, 144; green, 19; blue, 254 }  ,fill opacity=1 ][line width=0.75]    (98.13,75.45) -- (108.33,85.83) ;
\draw [fill={rgb, 255:red, 144; green, 19; blue, 254 }  ,fill opacity=1 ][line width=0.75]    (73.33,50.83) -- (83.83,61.33) ;
\draw [fill={rgb, 255:red, 144; green, 19; blue, 254 }  ,fill opacity=1 ][line width=0.75]    (109,50.83) -- (98.11,61.78) ;
\draw [fill={rgb, 255:red, 144; green, 19; blue, 254 }  ,fill opacity=1 ][line width=0.75]    (84,74.5) -- (73.33,85.83) ;

\draw  [fill={rgb, 255:red, 255; green, 255; blue, 255 }  ,fill opacity=1 ] (70.53,50.83) .. controls (70.53,49.29) and (71.79,48.03) .. (73.33,48.03) .. controls (74.88,48.03) and (76.13,49.29) .. (76.13,50.83) .. controls (76.13,52.38) and (74.88,53.63) .. (73.33,53.63) .. controls (71.79,53.63) and (70.53,52.38) .. (70.53,50.83) -- cycle ;
\draw [line width=1.5]    (63.5,43.94) -- (63.57,102.12) ;
\draw  [fill={rgb, 255:red, 208; green, 2; blue, 27 }  ,fill opacity=1 ] (70.89,82.33) .. controls (72.15,82.33) and (73.17,83.35) .. (73.17,84.61) -- (73.17,91.45) .. controls (73.17,92.71) and (72.15,93.73) .. (70.89,93.73) -- (56.18,93.73) .. controls (54.92,93.73) and (53.9,92.71) .. (53.9,91.45) -- (53.9,84.61) .. controls (53.9,83.35) and (54.92,82.33) .. (56.18,82.33) -- cycle ;
\draw  [line width=0.75]  (70.94,87.17) -- (70.94,91.17) -- (66.94,91.17) ;

\draw [line width=1.5]    (142.5,43.94) -- (142.57,102.12) ;
\draw  [fill={rgb, 255:red, 208; green, 2; blue, 27 }  ,fill opacity=1 ] (149.89,53.33) .. controls (151.15,53.33) and (152.17,54.35) .. (152.17,55.61) -- (152.17,62.45) .. controls (152.17,63.71) and (151.15,64.73) .. (149.89,64.73) -- (135.18,64.73) .. controls (133.92,64.73) and (132.9,63.71) .. (132.9,62.45) -- (132.9,55.61) .. controls (132.9,54.35) and (133.92,53.33) .. (135.18,53.33) -- cycle ;
\draw  [line width=0.75]  (149.94,58.17) -- (149.94,62.17) -- (145.94,62.17) ;

\draw    (152.5,58.13) -- (174,58.12) ;
\draw [fill={rgb, 255:red, 255; green, 255; blue, 255 }  ,fill opacity=1 ]   (152.33,85.83) -- (174.83,85.82) ;
\draw  [fill={rgb, 255:red, 255; green, 255; blue, 255 }  ,fill opacity=1 ] (149.53,85.83) .. controls (149.53,84.29) and (150.79,83.03) .. (152.33,83.03) .. controls (153.88,83.03) and (155.13,84.29) .. (155.13,85.83) .. controls (155.13,87.38) and (153.88,88.63) .. (152.33,88.63) .. controls (150.79,88.63) and (149.53,87.38) .. (149.53,85.83) -- cycle ;
\draw  [fill={rgb, 255:red, 144; green, 19; blue, 254 }  ,fill opacity=1 ][line width=1.5]  (269.83,63.79) .. controls (269.83,61.88) and (271.38,60.33) .. (273.29,60.33) -- (281.37,60.33) .. controls (283.28,60.33) and (284.83,61.88) .. (284.83,63.79) -- (284.83,71.87) .. controls (284.83,73.78) and (283.28,75.33) .. (281.37,75.33) -- (273.29,75.33) .. controls (271.38,75.33) and (269.83,73.78) .. (269.83,71.87) -- cycle ;
\draw  [fill={rgb, 255:red, 144; green, 19; blue, 254 }  ,fill opacity=1 ][line width=0.75]  (277.33,63.83) -- (281.33,63.83) -- (281.33,67.83) ;
\draw  [fill={rgb, 255:red, 255; green, 255; blue, 255 }  ,fill opacity=1 ] (294.2,46.83) .. controls (294.2,45.29) and (295.45,44.03) .. (297,44.03) .. controls (298.55,44.03) and (299.8,45.29) .. (299.8,46.83) .. controls (299.8,48.38) and (298.55,49.63) .. (297,49.63) .. controls (295.45,49.63) and (294.2,48.38) .. (294.2,46.83) -- cycle ;
\draw [line width=1.5]    (306.5,41.94) -- (306.57,101.12) ;
\draw  [fill={rgb, 255:red, 208; green, 2; blue, 27 }  ,fill opacity=1 ] (298.18,92.73) .. controls (296.92,92.73) and (295.9,91.71) .. (295.9,90.45) -- (295.9,83.61) .. controls (295.9,82.35) and (296.92,81.33) .. (298.18,81.33) -- (312.89,81.33) .. controls (314.15,81.33) and (315.17,82.35) .. (315.17,83.61) -- (315.17,90.45) .. controls (315.17,91.71) and (314.15,92.73) .. (312.89,92.73) -- cycle ;
\draw  [line width=0.75]  (298.13,87.9) -- (298.13,83.9) -- (302.13,83.9) ;

\draw [line width=1.5]    (367.6,102.05) -- (367.6,43.87) ;
\draw  [fill={rgb, 255:red, 208; green, 2; blue, 27 }  ,fill opacity=1 ] (360.15,62.71) .. controls (358.89,62.72) and (357.86,61.7) .. (357.85,60.44) -- (357.81,53.6) .. controls (357.8,52.34) and (358.81,51.32) .. (360.07,51.31) -- (374.78,51.21) .. controls (376.04,51.2) and (377.06,52.21) .. (377.07,53.47) -- (377.12,60.31) .. controls (377.13,61.57) and (376.12,62.6) .. (374.86,62.61) -- cycle ;
\draw  [line width=0.75]  (360.07,57.88) -- (360.04,53.88) -- (364.04,53.85) ;

\draw    (357.5,57.1) -- (337,57.09) ;
\draw [fill={rgb, 255:red, 255; green, 255; blue, 255 }  ,fill opacity=1 ]   (359.63,86.3) -- (337.13,86.31) ;
\draw  [fill={rgb, 255:red, 255; green, 255; blue, 255 }  ,fill opacity=1 ] (362.43,86.57) .. controls (362.42,88.12) and (361.15,89.36) .. (359.6,89.34) .. controls (358.06,89.32) and (356.82,88.06) .. (356.83,86.51) .. controls (356.85,84.96) and (358.12,83.72) .. (359.66,83.74) .. controls (361.21,83.76) and (362.45,85.02) .. (362.43,86.57) -- cycle ;
\draw    (281.37,75.33) .. controls (309.47,80.09) and (292.71,78.13) .. (281.71,81.13) ;
\draw    (273.29,75.33) .. controls (239.97,80.09) and (281.97,81.09) .. (295.47,87.09) ;
\draw    (258.54,86.29) -- (272.37,83.8) ;
\draw    (279.87,51.29) .. controls (296.2,53.63) and (299.2,56.29) .. (283.47,60.76) ;
\draw [fill={rgb, 255:red, 255; green, 255; blue, 255 }  ,fill opacity=1 ]   (270.54,48.33) -- (261.87,45.63) ;
\draw [line width=1.5]    (189,73.95) -- (242.44,74.45) ;
\draw [shift={(246.44,74.49)}, rotate = 180.54] [fill={rgb, 255:red, 0; green, 0; blue, 0 }  ][line width=0.08]  [draw opacity=0] (13.93,-6.69) -- (0,0) -- (13.93,6.69) -- cycle    ;

\draw (109,64.74) node [anchor=north west][inner sep=0.75pt]    {$=$};
\draw (315.8,65.14) node [anchor=north west][inner sep=0.75pt]    {$=$};

\end{tikzpicture}
\end{equation}

The reverse holds as well: a solution of the solvable condition with single-site MPS on the right can be mapped to the left solution by applying the {\footnotesize${\mathrm{SWAP}}$} gate. 

For $q=2,\tilde{q}=1$, we have exhausted the solutions fulfilling one-sided solvable condition as solved in Eq. \eqref{eq:TheSolution}:
\begin{equation}
    U_+=e^{i\phi}(u_+\otimes e^{-i\epsilon\sigma^3})V[\frac{\pi}{4},\frac{\pi}{4},J_3](e^{-i\eta\sigma^3}\otimes v_+).
\end{equation}
The subscript $+$ represents the chirality of the solvable condition.
Using the correspondence introduced above, we can directly present the solutions for the opposite-chirality solvable condition as:
\begin{equation}
    U_-=e^{i\phi}( e^{-i\epsilon'\sigma^3}\otimes u_-)V[\frac{\pi}{4},\frac{\pi}{4},J_3](v_-\otimes e^{-i\eta'\sigma^3}).
\end{equation}
Consequently, we can obtain the exhaustive parameterization for two-qubit gates fulfilling both solvable conditions by examining the overlap of two sets of solutions, which gives:
\begin{equation}
    U=e^{i\phi}(e^{-i\epsilon'\sigma^3}\otimes e^{-i\epsilon\sigma^3})V[\frac{\pi}{4},\frac{\pi}{4},J_3](e^{-i\eta\sigma^3}\otimes e^{-i\eta'\sigma^3}).
\end{equation}

For $q=4,\tilde{q}=2$, we have presented a subclass of solutions as Eq. (12) of the main text: $U_+=e^{i\phi}W_2SW_1(I\otimes v_+)$, where $v_+\in \text{SU}(4)$, $W_{1,2}=\sum_{a=0}^{3} f_{1,2}^{(a)}\otimes \ket{a}\bra{a}$, $f_1^{(a)}=
\left( \begin{array}{cc}
I_2 & 0 \\
0 & g^{(a)} 
\end{array} \right)$, $g^{(a)}\in\text{SU}(2)$, and $f_{2}^{(a)}\in \text{SU}(4)$ for all $a$. The corresponding opposite-chirality solutions can be parameterized as $U_-=e^{i\phi}SW_2SW_1S(v_-\otimes I)$. The overlap of these two sets of solutions gives
\begin{equation}
    U=e^{i\phi}(u_+\otimes u_-)S\exp(-iH)(v_-\otimes v_+),
\end{equation}
where $H$ is a real symmetric matrix $H\in \mathbb{R}^q\times \mathbb{R}^q$,  $H_{ab}=0$ for $a=0,1$ or $b=0,1$ , and $u_\pm, v_\pm$ take the direct sum form as $
\left( \begin{array}{cc}
I_2 & 0 \\
0 & w 
\end{array} \right)$, $w\in \text{SU}(2)$. 

\bibliographystyle{apsrev4-1-title}
\bibliography{Heran_circuit}